\documentclass[aps,nofootinbib,prb,reprint,superscriptaddress,longbibliography]{revtex4-1}
\usepackage{amsmath}
\usepackage{graphicx}
\usepackage{amsfonts}
\usepackage{color}
\usepackage{bm}
\usepackage{lipsum}
\usepackage{comment}
\usepackage{dsfont}
\usepackage{bm}
\usepackage{amsmath}
\usepackage{amssymb}
\usepackage{graphicx} 
\usepackage{braket}
\usepackage{mathtools}
\usepackage{times}
\usepackage{enumerate}
\usepackage{subcaption}

\usepackage[colorlinks,linkcolor=blue,urlcolor=blue,citecolor=blue]{hyperref}
\newcommand{\bea}{\begin{eqnarray}}
\newcommand{\eea}{\end{eqnarray}}
\usepackage[normalem]{ulem}
\usepackage{xcolor}
\newcommand{\redsout}[1]{\bgroup\markoverwith{\textcolor{red}{\rule[0.5ex]{2pt}{0.4pt}}}\ULon{#1}}

\begin{document}

\title{Compact cavity-dressed Hamiltonian framework at arbitrarily strong light-matter coupling}
%

\author{Jakub Garwo\l a}
\email{jakub.garwola@mail.utoronto.ca}

\affiliation{Department of Physics and Centre for Quantum Information and Quantum Control, University of Toronto, 60 Saint George St., Toronto, Ontario, M5S 1A7, Canada}

\author{Dvira Segal}
\email{dvira.segal@utoronto.ca}

\affiliation{Department of Physics and Centre for Quantum Information and Quantum Control, University of Toronto, 60 Saint George St., Toronto, Ontario, M5S 1A7, Canada}
\affiliation{Department of Chemistry, University of Toronto, 80 Saint George St., Toronto, Ontario, M5S 3H6, Canada}

\date{\today}

\begin{abstract}
We present a non-perturbative Hamiltonian mapping method for quantum systems strongly coupled to a quantized field mode (cavity), yielding compact closed-form representations of hybrid light–matter systems, allowing efficient studies of equilibrium and dynamical properties.
The mapping builds on an entangling transformation of photonic and atomic degrees of freedom. By truncating the resulting cavity-dressed Hamiltonian (CDH) to successively larger excitation sectors, we construct a series of compact models that converge to the exact limit, outperforming conventional approaches even in the challenging resonant and ultrastrong light–matter regime. 
We benchmark the CDH framework on the quantum Rabi model, obtaining accurate spectra and thermal observables across weak- and strong coupling regimes. We further apply the method to the {\it open} Rabi model, highlighting its computational advantages for non-unitary dynamics. As a nontrivial application, we analyze the Dicke–Heisenberg lattice model, determine its phase diagram under resonant and strong light–matter coupling, and reveal cavity-mediated spin correlations analytically and numerically, achieving substantial computational savings over brute-force simulations.
The closed-form and compactness of the CDH provide both physical insight and improved computational efficiency in studies of strongly coupled hybrid light–matter systems.
\end{abstract}
\maketitle

\textit{Introduction}---Under strong light–matter coupling, photons and material excitations hybridize into quasiparticles known as polaritons, with fundamentally transformed properties and behavior than in the bare material \cite{RMP-rev}.
Recent advances in the ultrastrong coupling regime enabled control over collective phenomena in quantum materials \cite{materials-rev}, realization of nonlinear optical effects and protection of quantum information \cite{Nori-rev}, modification of chemical reactivity \cite{Ebbesen23,Nitzan-rev,Pengfei23Rev}, and enhancement of transport processes in disordered media \cite{Transport-rev}.
%
Experiments and supporting calculations have exemplified, e.g., control of chemical reactivity through strong light-matter coupling to vibrational \cite{Ebbesen15,Ebbesen15b, Ebbesen16, Thomas16, MartinezMartinez18,Thomas19,Tal19,Hirai21,Ebbesen24, Li21,Cao21,Lindoy23,Bowan25}, electronic excitations \cite{Ebbesen12,conduc15,Tal18,Munkhbat18,stranius_selective_2018,Yuen19,Caonano22,Rubio23,Anael25,KeelingE25},
and engineered qubits \cite{wallraff_strong_2004,PRX15,Lupascu17,PRX18,Kuzmin19,Lupascu23}, with the strongest effect showing near resonance between molecular and cavity modes, a regime that remains theoretically challenging \cite{Yuen19C,Lindoy23,anto2023effective,Richardson23}. 
Beyond molecular systems, cavities have emerged as powerful platforms for materials engineering, where a tailored electromagnetic vacuum induces nonlinear optical properties \cite{mueller_deep_2020} and stabilizes quantum phases \cite{metal2insulator,Tureci12,Rubio23T}.

In the ultrastrong coupling regime (USC), light-matter
interaction energy becomes comparable to bare matter and light excitation frequencies, while in the {\it deep strong coupling} regime, it exceeds them \cite{RMP-rev}. 
Theoretically and computationally, describing the USC regime remains a challenge due to the breakdown of conventional perturbative treatments \cite{theory20}. 
Approaches built on, e.g., dressing transformations \cite{PhysRevLett.99.173601,anto2023effective,PhysRevA.108.043717,10.1063/1.5095940,Saller22,Saller23,NitzanR25}, variational ansatze \cite{peruzzo_variational_2014,Blais2020, Dave24}, 
Markovian embedding \cite{anto2023effective,garwola2024open,Garwola25,Feist21,Feist24},
asymptotic decoupling \cite{ashida2021}, 
perturbation theory on mean-field reference wavefunctions
\cite{Bauer23,pert25},
and generalized rotating-wave approximations \cite{James_Cummings,PhysRevLett.99.173601,PhysRevB.72.195410} have enabled significant progress in addressing the challenges of the USC regime.
However, theoretical results are limited to specific models (e.g., the Rabi model \cite{PhysRevLett.99.173601,Braak2011,Rabi17}), while 
computations offer limited insight into the nature of the USC regime, constraining the rational design of cavity-modified properties and dynamics.

In this letter, we present a general, compact, and analytically tractable framework for deriving cavity-dressed Hamiltonians (CDHs) describing quantum systems interacting with collections of bosonic modes at arbitrary coupling strengths, covering the weak, ultrastrong and deep-strong regimes. The approach is accurate, systematically convergent, and broadly applicable, encompassing both single-mode cavities and bosonic thermal environments \cite{SM}\nocite{Nazir2018,
anto2023effective,CURTRIGHT2015401,PhysRevA.108.043717,NitzanBook,PhysRevA.104.052617,Brett24,mendonca2025,hörmann2025commentrolematterinteractions,Zhang14,Schmidt2024,Schmidt2025,Kai25}, and remains accurate even in the resonant and ultrastrong light–matter coupling regime.
The main advantages of our method are:
(i) its analytical formulation, which provides physical insight and thus pathways to control light-induced matter processes;
(ii) its exactness in the deep strong coupling limit for coupled spin–boson systems; and
(iii) its straightforward numerical implementability.


We apply the CDH approach onto the quantum Rabi model \cite{PhysRev.49.324,Rabi17} and the Dicke–Heisenberg model \cite{PhysRev.93.99,heisenberg_zur_1928,mendonca2025,Zhang14,hörmann2025commentrolematterinteractions,grimaudo_landau-majorana-stuckelberg-zener_2019,grimaudo_spin-12_2017,holzinger_compact_2025,tong2025phasetransitionsopendicke,PhysRevLett.118.123602,PhysRevA.111.052415,PhysRevA.106.032212,Schmidt2020,Schmidt2024,Schmidt2025}, archetypal light-matter systems; it can be feasibly
applied to other cavity-coupled molecular and material Hamiltonians.
%
%
The Rabi model, which describes a two-level system coupled to a single bosonic mode, has attracted renewed attention with the advent of the ultrastrong and deep-strong coupling regimes in circuit quantum electrodynamics \cite{wallraff_strong_2004} and in emerging quantum simulation platforms \cite{braumuller_analog_2017}. Its integrability \cite{Braak2011} 
makes it an ideal benchmark for testing theoretical methods and resolving basic concepts in quantum optics \cite{ElinorSC22}.
The Dicke model extends the Rabi framework to a collection of spins coupled to a common bosonic mode \cite{Kirton19}.
Moreover, in the Dicke–Heisenberg model, a collection of spins 
in a chain configuration interact with their neighbors, as well as with a cavity mode, exhibiting cavity induced collective behavior 
\cite{mendonca2025}. 
%

\begin{figure}[hbpt]
\centering
\includegraphics[width=0.35\textwidth]{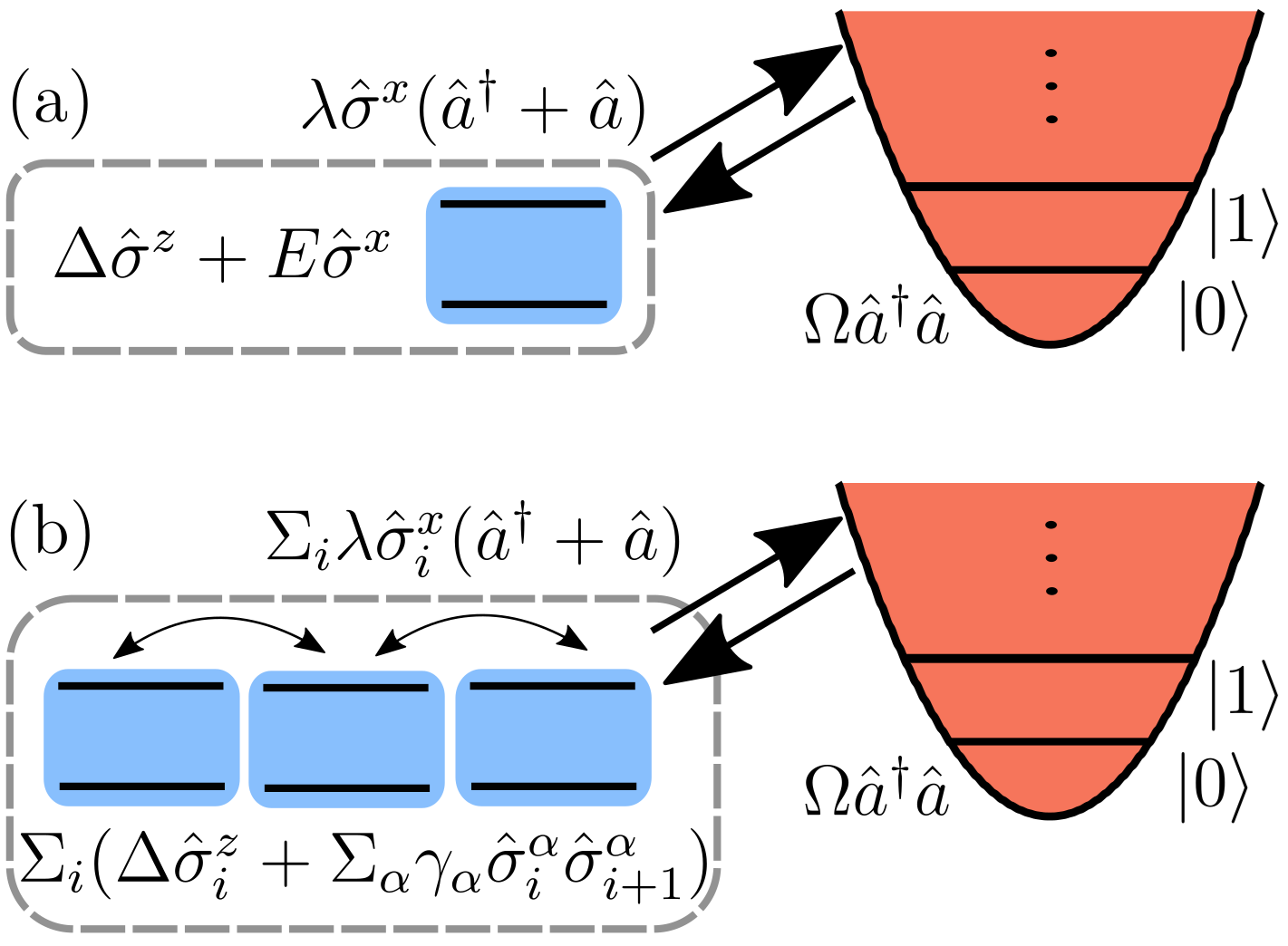}
\caption{ (a) The quantum Rabi model with a spin coupled to a single boson, and (b) the Dicke-Heisenberg model.}
\label{fig:Fig_1}
\end{figure}

\textit{Method}--- We consider a matter system, governed by the Hamiltonian $\hat{H}_S$, coupled to multiple bosonic modes, labelled $n$, via different system operators $\hat S_n$,
\begin{equation}
\hat{H}=\hat{H}_{S}+\sum_{n}\Big(\lambda_{n}\hat{S}_{n}(\hat{a}_{n}^{\dagger}+\hat{a}_{n})+\Omega_{n}\hat{a}_{n}^{\dagger}\hat{a}_{n}\Big).
\label{eq:H}
\end{equation}
Each bosonic mode (annihilation operator $\hat a_n$) is characterized by its frequency $\Omega_{n}$, and it couples to the system with strength $\lambda_{n}$. Fig. \ref{fig:Fig_1} depicts examples with a single lossless cavity mode. In Ref. \cite{SM} we generalize this to include  dissipation effects from secondary baths. 

The CDH mapping consists of a unitary light-matter entangling transformation,  $\hat{H}\rightarrow\hat{U}_P\hat{H}\hat{U}_P^\dagger$ with $ \hat{U}_{P} =\exp [\sum_{n} \frac{\lambda_{n}} { \Omega_{n}} \hat{S}_{n}(\hat{a}_{n}^{\dagger}-\hat{a}_{n})]$ the familiar polaron transform, generalized to the multi-mode case \cite{garwola2024open}. 
The total Hilbert space is then decomposed into blocks defined by the occupation number of modes, $( \hat{H} )^{\text{cdh}}_{\boldsymbol{s}\boldsymbol{r}}=\langle\boldsymbol{s}|\hat{U}_P\hat{H}\hat{U}_P^\dagger|\boldsymbol{r}\rangle$.
For example, $|\boldsymbol{s}\rangle =|0 0\rangle$ corresponds to two bosonic modes in their ground states. 
%
Each block is an operator on the system Hilbert space. To evaluate these blocks, we use a momentum representation for each mode and turn the expectation value calculation into an integral over momenta. This approach is particularly powerful when coupling operators $\hat{S}_n$ do not commute \cite{garwola2024open}. We truncate each  mode to its $M$ lowest energy states in the basis of particle numbers. The result is a cavity-dressed Hamiltonian,
\begin{equation}
\begin{aligned}&\hat{H}_{(M)}^{\text{cdh}}  =(\hat{H}_S)_{(M)}^{\text{cdh}}+\sum_n \left[ \Omega_n\hat{N}_{n,(M)}-\frac{\lambda^{2}_n}{\Omega_n} \left(\hat{S}^2_{n} \right)^\text{cdh}_{(M)} \right],
\label{eq:H_EFF_M}
\end{aligned}
\end{equation}
where $(\hat{\mathcal{O}})_{(M)}^{\text{cdh}}=\sum_{\boldsymbol{s},\boldsymbol{r}}|\boldsymbol{s}\rangle\langle\boldsymbol{r}|\otimes(\hat{\mathcal{O}})_{\boldsymbol{s}\boldsymbol{r}}^{\text{cdh}}$ with $(M)$ indicating the degree of truncation. For details, see Ref.  \citenum{SM}. $\hat{N}$ is the number operator. To illustrate the applicability of the CDH, in what follows we focus on models with a single cavity mode.

To compute expectation values in the CDH framework, system observables in the lab frame are written as \cite{10.1063/5.0228779}
$\langle \hat{\mathcal{O}} \rangle_\rho = \text{Tr}( \hat{\mathcal{O}}  \rho ) = \text{Tr} ( \hat{U}_P \hat{\mathcal{O}} \hat{U}_P^\dagger \rho_P )$, with $\rho_P$ the density matrix in the polaron frame. We compute $\rho_P$ assuming a canonical state, taking the CDH as the mean-force Hamiltonian \cite{MFGS22,SM}.
Rotated operators must also be truncated: If $\hat{\mathcal{O}}$ does not commute with $\hat{U}_P$, we rotate it including $M_P$ levels per mode, then truncate to dimension $M$.
Setting $M_{P}=M$ offers the greatest computational efficiency.
In certain cases \cite{SM}, using  $M_{P}>M$ to perform the rotation, before truncating back to dimension $M$ was necessary, 
but never exceeding the dimension $N$ needed for convergence in the conventional-bare approach.
Alternatively, for a single cavity mode, or multiple modes with commuting coupling operators, matrix elements of $\hat{U}_P \hat{\mathcal{O}} \hat{U}_P^\dagger$ are found exactly by performing the spectral decomposition of $\hat{S}_n$ \cite{SM}. 


\textit{Quantum Rabi Model}---We illustrate the CDH representation on the quantum Rabi model. After analytically deriving the degree-$M$ CDH, we compute the model's energy spectrum, thermal equilibrium, and dissipative behavior. 
Comparing these results with exact analytical and numerical calculations, we show the superior convergence of the CDH framework relative to the bare representation, as well as the physical insights provided by its block structure. 
The model is described by the following Hamiltonian $\hat{H}=\Delta \hat{\sigma}^z + 
    \Omega \hat{a}^\dagger \hat{a} + \lambda \hat{\sigma}^x \left( \hat{a}^\dagger + \hat{a} \right)$, with $2\Delta$ the spin splitting, $\Omega$  the cavity frequency and $\lambda$ as the spin-cavity coupling strength. The model is integrable; the analytical solution for its eigenenergies is given as the roots of an infinite sum of recursively defined real functions \cite{Braak2011}. When simulations are performed in the bare representation, we denote the dimension of the cavity Hilbert space by $N$; for CDH simulations, we use $M$.

The CDH Hamiltonian for the Rabi model is derived in Ref.~\cite{SM}. For $M=1$, we get  $\hat{H}_{(1)}^{\text{cdh}} = \tilde{\Delta}\hat{\sigma}^{z} - \Omega\epsilon^2 \hat I$, which corresponds, in the polaron frame, to a spin coupled to the cavity ground state.
Here, $\epsilon=\lambda/\Omega$ and $\tilde{\Delta}=\Delta e^{-2\epsilon^2}$ is the suppressed spin splitting. 
For $M = 2$, the cavity is truncated to two levels, and the  CDH takes the form
\begin{equation}
\begin{aligned}\hat{H}_{(2)}^{\text{cdh}} & =\begin{pmatrix}\tilde{\Delta}\hat{\sigma}^{z}-\Omega\epsilon^{2}\hat{1} & 2\tilde{\Delta}\epsilon i\hat{\sigma}^{y}\\
-2\tilde{\Delta}\epsilon i\hat{\sigma}^{y} & \tilde{\Delta}(1-4\epsilon^{2})\hat{\sigma}^{z}+\Omega(1-\epsilon^{2})\hat{1}
\end{pmatrix}.
\label{eq:H_EFF_2_rabi}
\end{aligned}
\end{equation}
The eigenenergies of $\hat{H}_{(2)}^{\text{cdh}}$ are
\begin{equation}
\begin{aligned}
&E_{\pm,\pm'}=\frac{1}{2}\Omega-\epsilon^{2}(\Omega\pm2\tilde{\Delta})\\ &\pm'\sqrt{\tilde{\Delta}^{2}\left(4\epsilon^{4}+1\right)\pm\tilde{\Delta}\Omega\left(1-2\epsilon^{2}\right)+\Omega^{2}/4}.
\label{eq:Hcdh2}
\end{aligned}
\end{equation}
The analytical form of the CDH provides insight into the resulting hybrid states. 
In the weak-coupling limit, $\epsilon \to 0$, the spectrum reduces to that of a decoupled spin and a truncated free cavity. In the opposite, deep-strong limit and for any truncation degree $M$, the CDH retains only the terms $\Omega\hat{N}_{(M)}-\Omega\epsilon^2\hat{1}_{(M)}$ on diagonal blocks, which correspond to the cavity excitation ladder and a global energy shift  induced by the spin–cavity coupling. This universal behavior, independent of $M$, explains why even for $M = 1$ the CDH yields accurate predictions for the Rabi model (as well as the Dicke-Heisenberg model below) in the $\lambda \to \infty$ limit \cite{anto2023effective}.
Between the ultraweak and deep-strong coupling limits, the off-diagonal terms in Eq.~(\ref{eq:Hcdh2}), as well as in higher-order CDHs \cite{SM}, reveal that transitions between cavity levels $l \to l \pm p$ are coupled to system transitions with an effective strength
$\frac{2\Delta}{\Omega}\times \left(\frac{\lambda}{\Omega}\right)^{p-1}  \lambda e^{-2\lambda^2/\Omega^2}$.
It is thus clear that the resonance condition ($2\Delta = \Omega$) is more favorable for applications than the off-resonant case ($2\Delta < \Omega$), as it effectively enhances light-matter interactions. 

\begin{figure*}[hbpt]
\centering
\includegraphics[width=0.95\textwidth]{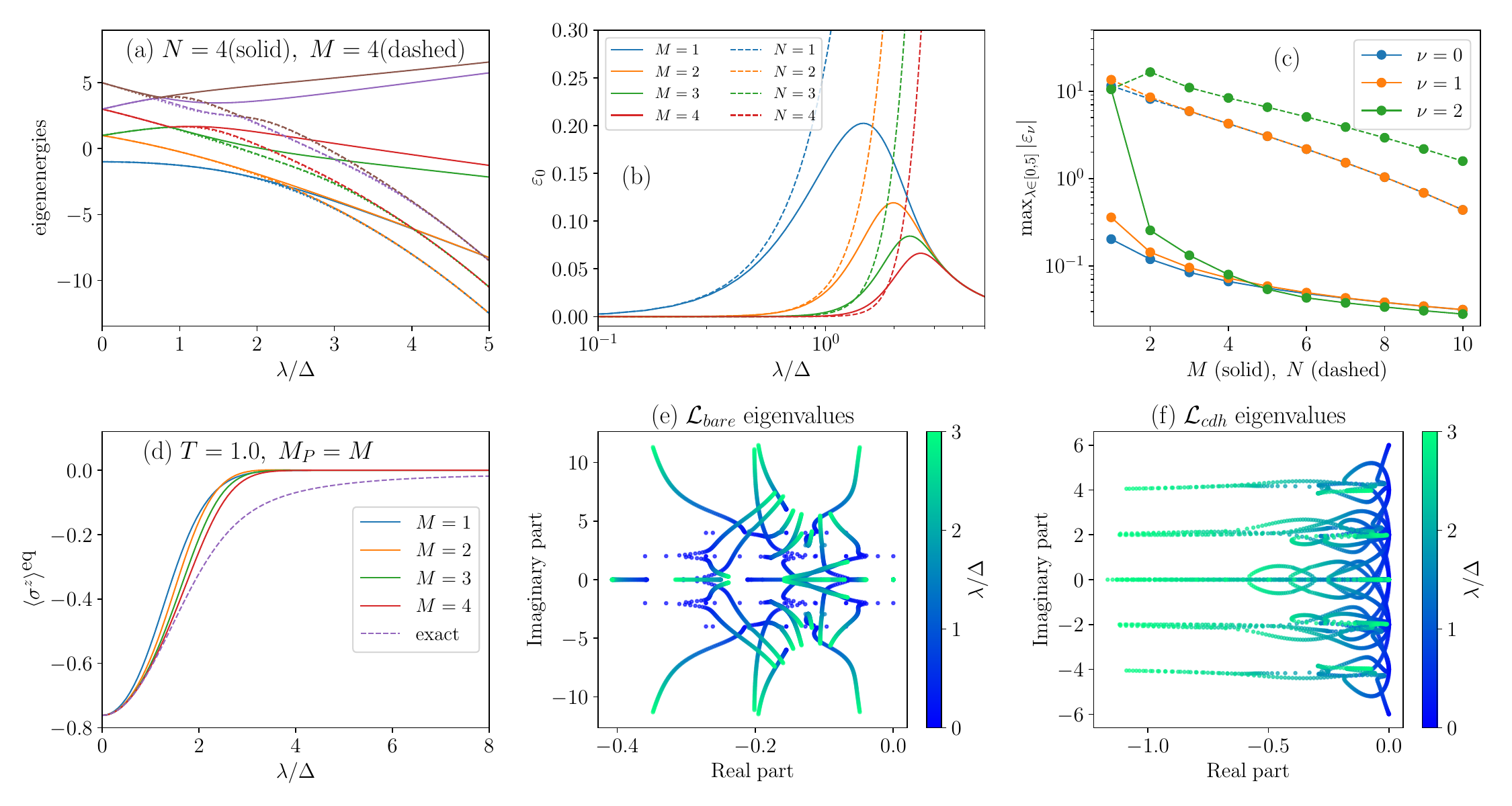}
\caption{(a) Eigenenergies of the quantum Rabi model against the coupling strength $\lambda$ using the bare representation with $N=4$ (solid) and a CDH of degree $M=4$ (dashed). Exact eigenvalues\cite{Braak2011} (dotted) overlap with CDH results. 
(b) $\varepsilon_0$ defined as the difference between the zero eigenenergy predicted by the CDH with $M\in[1,4]$ or the bare model with $N\in[1,4]$, and the exact result.
(c) Maximal error $\varepsilon_\nu$ over $\lambda\in[0,5]$ with respect to $M$ and $N$ for the first three eigenenergies, $\nu=0,1,2$.
(d) Equilibrium magnetization $\langle \hat{\sigma}^z \rangle_{eq}$ at $T=1$ using the CDH with $M\in[1,4]$, compared to brute force converging solution (dashed). Parameters are $\Omega=2$, $\Delta=1$. 
(e)-(f) Complex eigenvalues of the  Redfield generator in the bare 
($N=3$) and the CDH 
($M$=3) representations. For the dissipator we assume an ohmic function with $\gamma=0.01$ as a dimensionless cavity-bath coupling coefficient, high frequency cutoff, and $T=1$.
Colors correspond to different values of $\lambda$.  At this truncation, the generator has $(2\times 3)^2$  eigenvalues for each $\lambda$.
}
\label{fig:Rabi}
\end{figure*}

In Fig. \ref{fig:Rabi}(a) we analyze the first six eigenenergies of the quantum Rabi model as a function of the coupling strength in the resonant case. 
We present exact results (dotted, overlapped with dashed) \cite{Braak2011}, and compare them with numerical simulations using the CDH of degree $M=4$ (dashed)
and bare-conventional simulations with $N = 4$ cavity levels (full).
In bare simulations with $N = 4$ cavity levels, the ground-state energy diverges beyond $\lambda / \Delta \approx 2$ [see Fig.~\ref{fig:Rabi}(b)], with excited-state energies exhibiting even larger deviations, Fig.~\ref{fig:Rabi}(c).
%
In contrast, the CDH accurately reconstructs the first six energy levels in all coupling strengths, maintaining a {\it bounded error} [Fig.~\ref{fig:Rabi}(b)]
in both weak-, ultrastrong-, and deep-coupling limits; the residual error in the CDH arises primarily in the coupling regime around $\lambda / \Delta \in [1, 5]$. This represents a substantial reduction in computational cost for numerical simulations.
%
%
In Fig. \ref{fig:Rabi}(c), we study the maximum of the absolute error $\varepsilon_\nu$ over $\lambda\in[0,5]$ for the first three energy levels using the CDH method ($M$, solid) and the bare truncation ($N$, dashed). With CDH mapping, we observe a striking improvement in error scaling with Hilbert-space dimension, particularly for small $M$. Notably, the CDH mapping exhibits comparable error scaling for excited states, whereas the bare-basis calculation becomes progressively less accurate, beginning with the second excited-state energy.

Having established the CDH representation's convergence of eigenenergies across all coupling strengths, we use it to compute 
thermal equilibrium observables. 
In Fig.~\ref{fig:Rabi}(d), we present the equilibrium magnetization as a function of $\lambda$, $\langle \hat{\sigma}^z \rangle^{\mathrm{eq}} = \frac{1}{Z}\text{Tr}[ \hat{\sigma}^z e^{- \hat{H}^{\text{cdh}}_{(M)}/T}]$; $Z=\text{Tr}[e^{-\hat{H}^{\text{cdh}}_{(M)}/T}] $ is the canonical partition function, 
 $k_B=1$. 
CDH simulations are compared to results from the 
exact solution \cite{Braak2011}. It is evident that the series of CDHs yields progressively convergent observables as we increase $M$, not only in the weak- and ultrastrong-coupling limits, but also in the intermediate regime. 
%
%
In Ref.~\cite{SM}, we examine the average magnetization at both zero temperature (ground state) and finite temperature (thermal state). For ground state observables, $M_{P}>M$ was required to achieve convergence. We also study there the cavity occupation number, demonstrating the advantage of CDH in convergence over the bare approach, up to the ultrastrong coupling regime.

We now use the CDH formalism to study dissipative open system dynamics, 
described by a quantum master equation, $\dot{\rho}= \mathcal{L}\rho$, where here we choose the dynamics generator $\mathcal{L}$ to be of the Redfield form assuming weak dissipation \cite{NitzanBook}. In our model, the cavity is coupled to a dissipative thermal bath;
the microscopic description is given in Ref. \citenum{SM}.
%
We characterize the open-system dynamics and its relaxation toward the steady state through the spectrum of the Liouvillian $\mathcal{L}$. In the doubled Hilbert-space representation, the Liouvillian eigenvalue equation reads
$
\hat{\mathcal{L}} \ket{\!\ket{\rho_n}} = \left(-\Gamma_n^{\mathrm{R}} + i\,\Gamma_n^{\mathrm{I}}\right)\ket{\!\ket{\rho_n}}$,
where the real and imaginary parts of the eigenvalues, $\Gamma_n^{\mathrm{R}}$ and $\Gamma_n^{\mathrm{I}}$, correspond, respectively, to the decay rate and oscillation frequency of the eigenmode $\ket{\!\ket{\rho_n}}$. The index $n$ runs over the full Liouville space, 
$n \in \{ 1, \dots, \text{dim}(\hat{H})^2 \}$.

Figure \ref{fig:Rabi}(e)–(f) illustrates the spectrum of the Redfield generator for the open Rabi model. 
In the bare representation, the spectrum fails to converge for $N=3$. In contrast, the CDH spectrum ($M=3$) is convergent and it exhibits a strikingly organized structure: The eigenvalues separate into distinct branches, vertically spaced by $\Omega$, and shifted horizontally at large $\lambda$ as the coupling is increased. 
These branches correspond to different oscillatory modes in the dissipative dynamics. Within each branch, as we increase $\lambda$, at least one eigenvalue remains close to the $\Gamma^{\mathrm{R}}=0$ axis (slow dissipative decay), while another diverges toward $-\infty$ (fast decay). This reveals that a separation of timescales occurs at strong coupling for each frequency component. In Ref.~\cite{SM}, we illustrate these features in the magnetization dynamics of the open Rabi model. 
The CDH framework thus allows efficient-accurate calculations of {\it open} cavity-coupled systems.
%
%

\begin{figure*}[hbpt]
\centering
\includegraphics[width=1.0\textwidth]{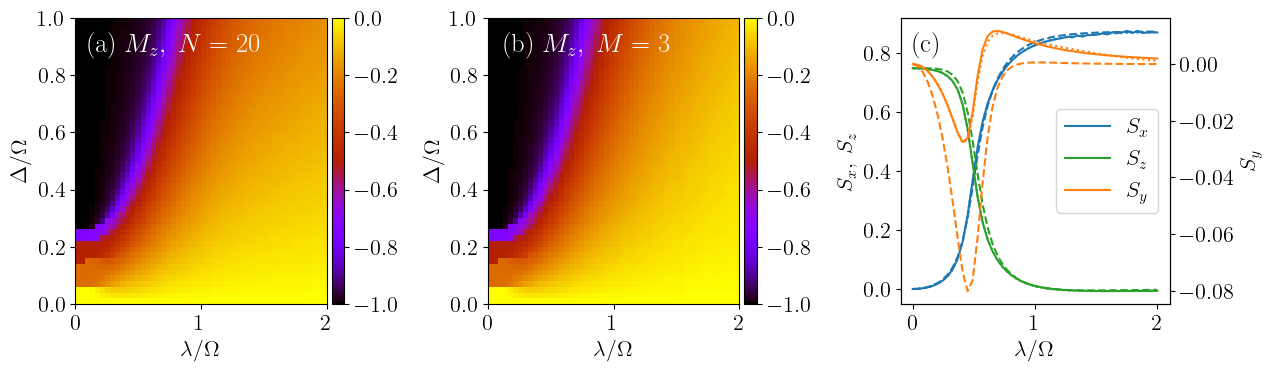}
\caption{Ground state phase diagrams of the Dicke-XX Heisenberg model with periodic boundary conditions (a) in the bare representation with $N=20$, (b) and using the $M=3$ CDH. The order parameter is the average magnetization, $M_z$. (c) Structure factors $S_\alpha$ for a resonance situation, $2\Delta=\Omega$. We compare numerically-converged simulations at $N=20$ (solid, this choice ensured that the maximal absolute error in magnetization remained below $0.01$) to the CDH with $M=1$ (dashed) and $M=3$ (dotted), with $M_P=M$. Fixed parameters are $L=8$, 
$\gamma_x=\gamma_y=\Omega/8$, $\gamma_z=0$, $\Omega=2$, and we vary $\lambda$ and $\Delta$.
}
\label{fig:phase_diagram}
\end{figure*}


\textit{Dicke-Heisenberg model}--- This model includes a Heisenberg spin chain coupled to a single-mode cavity,\cite{grimaudo_landau-majorana-stuckelberg-zener_2019,grimaudo_spin-12_2017}. We show that the CDH framework provides analytical insights and efficient simulations into the magnetic phases of the spin system in the Dicke-Heisenberg model.
For two sites, the model can be analytically solved by mapping it to the Rabi model \cite{Grimaudo2023}. Here, we consider an $L$-long chain,
\begin{equation}
\begin{aligned}\hat{H} & =\Delta\sum_{i=1}^{L}\hat{\sigma}_{i}^{z}+\Omega\hat{a}^{\dagger}\hat{a}\\
+ & \sum_{i=1}^{L}\left(\frac{\lambda}{\sqrt{L}}\hat{\sigma}_{i}^{x}(\hat{a}^{\dagger}+\hat{a})-\sum_{\alpha=x,y,z}\gamma_{\alpha}\hat{\sigma}_{i}^{\alpha}\hat{\sigma}_{i+1}^{\alpha}\right).
\label{eq:H_DH}
\end{aligned}
\end{equation}
$2\Delta$ is the spin splitting, $\Omega$ is the cavity frequency, $\lambda$ is the interaction strength, and $\gamma_{\alpha}$ are the Heisenberg interaction parameters. This model was studied numerically in the context of superradiant phenomena \cite{holzinger_compact_2025,tong2025phasetransitionsopendicke,PhysRevLett.118.123602} in many-body systems \cite{mendonca2025,PhysRevA.111.052415}, and for the design of quantum batteries \cite{PhysRevA.106.032212}. Here, we focus on uncovering features of the spin system in weak-to the deep-strong limit {\it using a small} $M$. To build the CDH, we transform this Hamiltonian with the unitary $ \hat{U}=\exp[ \; \frac{\lambda}{\Omega\sqrt{L}}\sum_{i=1}^{L}\hat{\sigma}_{i}^{x}(\hat{a}^{\dagger}-\hat{a})\; ]$ and truncate it to include $M$ excitation sectors \cite{SM}
\bea
\hat{H}^{\text{cdh}}_{(M)}  &=&\Delta\sum_{i=1}^{L}\left(\hat{\sigma}_{i}^{z}\right)_{(M)}^{\text{cdh}}-\frac{\lambda^{2}}{\Omega L}\sum_{i,j=1}^L\left( \hat{\sigma}^x_i \hat{\sigma}^x_j \right)^{\text{cdh}}_{(M)}+\Omega\hat{N}_{(M)}
 \nonumber\\
 & -&\sum_{i=1}^{L}\sum_{\alpha=x,y,z}\gamma_{\alpha}\left(\hat{\sigma}_{i}^{\alpha}\hat{\sigma}_{i+1}^{\alpha}\right)_{(M)}^{\text{cdh}}.\label{eq:H_DH_EFF}
\eea
The mapped one-body terms are identical to those in the Rabi model. The second term in the first line describes a cavity-mediated all-to-all spin coupling in the $x$ orientation. The second line includes Heisenberg two-body terms, modified due to the cavity coupling. In this group, the $\hat{\sigma}_{i}^{x}\hat{\sigma}_{j}^{x}$  terms remain unchanged under the entangling transformation, but the other two interaction terms transform in a non-trivial way: On diagonal blocks (corresponding to a fixed cavity occupation number $M$), they undergo mixing,
\begin{equation}
\begin{aligned}\hat{\sigma}_{i}^{y}\hat{\sigma}_{i+1}^{y} & \rightarrow f_{M}(\epsilon)\hat{\sigma}_{i}^{y}\hat{\sigma}_{i+1}^{y}+g_{M}(\epsilon)\hat{\sigma}_{i}^{z}\hat{\sigma}_{i+1}^{z},\\
\hat{\sigma}_{i}^{z}\hat{\sigma}_{i+1}^{z} & \rightarrow g_{M}(\epsilon)\hat{\sigma}_{i}^{y}\hat{\sigma}_{i+1}^{y}+f_{M}(\epsilon)\hat{\sigma}_{i}^{z}\hat{\sigma}_{i+1}^{z},
\label{eq:heisenberg_mixing}
\end{aligned}
\end{equation}
where $\epsilon=\lambda/(\Omega \sqrt{L})$ and $f_{M},g_{M}$ are dressing functions of the two-body operators in the $M$th diagonal block \cite{SM}. Intriguingly, in off-diagonal blocks, corresponding to transitions in the cavity, terms proportional to $\left(\gamma_{y}-\gamma_{z}\right)\left(\hat{\sigma}_{i}^{y}\hat{\sigma}_{i+1}^{z}+\hat{\sigma}_{i}^{z}\hat{\sigma}_{i+1}^{y}\right)$ appear. These terms {\it disappear} for the isotropic Heisenberg interaction. In Ref. \citenum{SM}, we provide closed-form expressions for the $M=3$ CDH. We note that in the deep strong coupling limit, the one body terms are exponentially suppressed, the Heisenberg interaction energies saturate to a constant, and the all-to-all interaction in the $x$ orientation scales as $\frac{\lambda^2}{\Omega}$. To understand the interplay between the different terms in Eq. (\ref{eq:H_DH_EFF}) we examine the phase diagram of the Dicke-Heisenberg model using two order parameters: ground state average magnetization, $M_z=\frac{1}{L}\sum_{i}\langle \hat\sigma_i^z\rangle_{GS}$ and ground state entanglement entropy $S=-\text{Tr}(\rho_A\ln\rho_A)$, where the subsystem $A$ consists of the first $L/2$ spins (assuming an even number of spins). 

In Fig. \ref{fig:phase_diagram}(a)-(b) we present the ground state phase diagram of the Dicke-XX Heisenberg model based on the magnetization; a similar crossover is captured with the entanglement entropy \cite{SM}.
In the bare representation, at weak cavity coupling, we observe an abrupt transition from a ferromagnetic to a paramagnetic phase as we reduce $\Delta$ and reach $\Delta/\Omega \approx 0.25$, corresponding to $\Delta \approx \gamma_x + \gamma_y$. This transition is already accurately captured by the CDH of order $M = 1$ \cite{SM}. Physically, when $\Delta > \gamma_x + \gamma_y$, the single-spin term dominates, resulting in a fully polarized ground state with $M_z = -1$ and vanishing entanglement entropy. 
Conversely, when $\Delta < \gamma_x + \gamma_y$, the Heisenberg interaction dominates, entangling spins into singlet-like pairs and driving $M_z$ to $0$ with $S>0$. In contrast, 
 the phase diagram of the isotropic Dicke-XXX Heisenberg model 
does not show such a transition at weak coupling \cite{SM}, as the $SU(2)$ symmetry of the spin interaction prevents the spins from aligning on the $x-y$ plane. 

%

Turning on the cavity coupling, we observe a continuous suppression of magnetization as the spin splitting $\Delta$ is reduced or the cavity coupling $\lambda$ increases. This behavior can be understood in terms of a cavity-mediated all-to-all interaction between spins along the $x$ direction. At strong coupling, this interaction dominates, generating spin entanglement and reducing magnetization along the $z$ axis.
%
In Fig. \ref{fig:phase_diagram}(c) we plot the structure factors $S_\alpha=\sum_{i,j} \langle \hat{\sigma}^{\alpha}_i \hat{\sigma}^{\alpha}_j \rangle_{GS}/L^2$, where $\alpha\in \{ x,y,z \}$, against the renormalized coupling $\lambda/\Omega$. We observe a monotonic increase of $S_x$ with $\lambda$, driven by cavity-mediated all-to-all interactions, suppression of single-spin splitting terms, and renormalization of the XX Heisenberg interaction. Correspondingly, $S_z$ exhibits the opposite trend. The behavior of $S_y$, on the other hand, is nonmonotonic, reflecting the mixing of spin terms, described in Eq.~(\ref{eq:heisenberg_mixing}).
We note that the structure factor readily converges to the numerically-converged result (full) as we increase $M$. In fact, for $S_{x,z}$, results are quantitatively correct for $M=1$, while for $S_y$, a CDH of degree $M=3$ is quantitatively correct.


\textit{Discussion}--- We introduced a non-perturbative and computationally efficient mapped Hamiltonian representation for simulating 
the structure, equilibrium behavior and relaxation dynamics of cavity-coupled quantum systems. The approach favorably converges in both resonant and off resonant regimes, and at all coupling strength. The technique combines a polaron-like transformation with a truncation of the cavity to $M$ levels. The resulting {\it closed-form} cavity-dressed Hamiltonians provide insights into light–matter hybrid effects. 
We tested the framework on the quantum Rabi impurity model and the Dicke–Heisenberg lattice model. 
%
The CDH captures essential features of the phase diagram of the  Dicke–Heisenberg model with a low-degree truncation, offering a substantial computational advantage over conventional tools. 
The method can be applied to multimode cavities, including cases where modes couple to the system through distinct, noncommuting system operators \cite{SM,garwola2024open,Garwola25}. The CDH can be integrated with other open system formalisms to efficiently describe relaxation dynamics of strongly coupled hybrid light-matter systems, from the weak to the ultrastrong dissipative regimes.


\textit{Acknowledgements}--- We acknowledge useful discussions with Brett Min and Zachary Giles. JG is supported through the Ontario Graduate Scholarship, the Lachlan Gilchrist Fellowship, and the research project: ``Quantum Software Consortium: Exploring Distributed Quantum Solutions for Canada" (QSC). QSC is financed under the National Sciences and Engineering Research Council of Canada (NSERC) Alliance Consortia Quantum Grants \#ALLRP587590-23. D.S. acknowledges the NSERC Discovery Grant.

\textit{Data availability}--- All data presented in this manuscript is available in the referenced GitHub repository \cite{garwola_github}.

\begin{widetext}

{\center {\Large \bf Supplementary Material\\}} 
\vspace{10mm}
\noindent { \bf \large Title:}
{\large Compact cavity-dressed Hamiltonian framework of spin systems at arbitrarily strong coupling} \\

\noindent {\large \bf  Authors:} {\large Jakub Garwo\l a and Dvira Segal }\\

\vspace{5mm}
This Supplementary Material includes in Sec. \ref{sec:AppA} details on the derivation of the cavity-dressed Hamiltonians (CDH). The derivation for general multi-mode cavity configurations is presented in Sec. \ref{sec:Amodes}, followed by a generalization to include coupling to a secondary dissipative bath in Sec. \ref{sec:Abaths}.
We discuss considerations in the CDH simulations in \ref{sec:Anum}.
In Sec. \ref{sec:AppB} we apply the mapping onto the quantum Rabi model presenting the closed-form $M=4$ CDH. Smaller CDH are given by leaving out higher-order blocks. We further complement results from the main text with additional simulations.
We derive the CDH for the Dicke-Heisenberg model in Sec. \ref{sec:AppC}, providing the $M=3$ CDH of the model. We include  additional results for the Dicke-XX model, as well as simulations of other models: The isotropic Dicke-Heisenberg model, and the Dicke-Ising model.

To organize the notation, in Table \ref{tab:definitions} we include a summary of the most important definitions used in this manuscript.

\begin{table}[htbp]
\centering
\caption{Definitions of symbols and parameters used in the text.}
\label{tab:definitions}
\begin{tabular}{ll}
\hline\hline
\textbf{Symbol} & \textbf{Definition} \\
\hline
$\hat{H}$ & Total light--matter Hamiltonian \\
$\hat{H}_S$ & Bare matter (system) Hamiltonian \\
$\hat{S}_n$ & System operator coupling to bosonic mode $n$ \\
$\hat{U}_P$ & (Mixed) polaron unitary transformation \\
$\rho$ & Density matrix in the laboratory (bare) frame \\
$\rho_P$ & Density matrix in the polaron frame \\
$\hat{N}_n$ & Number operator of bosonic mode $n$ \\
$\boldsymbol{r},\boldsymbol{s}$ & Multi-indices labeling bosonic occupation-number sectors \\
\hline
$M$ & Truncation dimension of the cavity Hilbert space in the CDH \\
$N$ & Truncation dimension of the cavity Hilbert space in the bare representation \\
$M_P$ & Number of bosonic levels retained when rotating operators by $\hat{U}_P$ \\
\hline
$\hat{H}^{\mathrm{cdh}}_{(M)}$ & Degree-$M$ cavity-dressed Hamiltonian (CDH) \\
$(\hat{\mathcal O})^{\mathrm{cdh}}_{(M)}$ & Operator $\hat{\mathcal O}$ mapped into the degree-$M$ CDH \\
\hline
$\Delta$ & Spin (two-level system) energy splitting \\
$\Omega$ & Single cavity-mode frequency \\
$\lambda$ & Spin--cavity coupling strength (single-mode case) \\
$\epsilon$ & Dimensionless coupling, $\epsilon=\lambda/\Omega$ \\
$\tilde{\Delta}$ & Renormalized spin splitting, $\tilde{\Delta}=\Delta e^{-2\epsilon^2}$ \\
$\hat{\sigma}^\alpha$ & Pauli matrices, $\alpha\in\{x,y,z\}$ \\
\hline
$L$ & Number of spins (sites) in the Dicke--Heisenberg chain \\
$\gamma_\alpha$ & Heisenberg interaction strength along direction $\alpha$ \\
$f_M(\epsilon),\,g_M(\epsilon)$ & Dressing functions for two-body spin operators in the diagonal CDH sectors \\
\hline
$Z$ & Canonical partition function \\
$T$ & Temperature \\
$M_z$ & Average magnetization, $M_z=\frac{1}{L}\sum_i\langle\hat{\sigma}_i^z\rangle$ \\
$S$ & Bipartite entanglement entropy of the spin subsystem \\
$S_\alpha$ & Spin structure factor along direction $\alpha$ \\ \hline

$\hat{H}_O$ & Hamiltonian of a light-matter interacting system, coupled to a bosonic bath \\

$J(\omega)$ & Bath spectral density function \\

$\hat{\mathcal{L}}$ & Dynamics generator, hat represents doubled Hilbert space representation \\

$-\Gamma^R_n+i\Gamma_n^I$ & Complex eigenvalue of the dynamics generator \\

\hline\hline
\end{tabular}
\end{table}

\section{Derivation of the CDH}
\label{sec:AppA}

\subsection{Multi-mode mapping}
\label{sec:Amodes}

In this Section, we derive the general form of the CDH
as presented in  Eq. (2) in the main text.
At this stage, we do not specify the system, nor its coupling operators to the cavity, thus the derivation is completely general in this respect. The mapped CDH can be used to study light-matter models with matter represented, e.g., by an impurity system, such as the Rabi model, 
an ensemble of subsystems, such as in the Dicke model, and lattices such as the Dicke-Heisenberg model and its many variants.

We consider a system coupled to multiple cavity modes,
\begin{equation}
\hat{H}=\hat{H}_{S}+\sum_{n}\Big(\lambda_{n}\hat{S}_{n}(\hat{a}_{n}^{\dagger}+\hat{a}_{n})+\Omega_{n}\hat{a}_{n}^{\dagger}\hat{a}_{n}\Big).
\end{equation}
Here $n$ is an index for bosonic modes, representing cavity modes. Each mode 
(annihilation operator $\hat{a}_n$) is characterized by a coupling strength $\lambda_n$, frequency $\Omega_n$. The system is described by its Hamiltonian $\hat{H}_S$ and a coupling operator to the $n$-th cavity mode $\hat{S}_n$. Later in this section, we consider a generalization of this formalism to leaky cavities. 
The CDH mapping method consists of a unitary transformation, $\hat{H}\rightarrow\hat{U}_P\hat{H}\hat{U}_P^\dagger$ and a decomposition of the total Hilbert space into subspaces 
 corresponding to occupation numbers of each mode. If $N_b$ is the number of modes, 
 we decompose the transformed Hamiltonian into blocks
\begin{equation}
\hat{U}_P\hat{H}\hat{U}^\dagger_P=\sum_{\boldsymbol{r},\boldsymbol{s}}|\boldsymbol{s}\rangle\langle\boldsymbol{s}|\hat{U}_P\hat{H}\hat{U}^\dagger_P|\boldsymbol{r}\rangle\langle\boldsymbol{r}|,
\end{equation}
where $\boldsymbol{r},\boldsymbol{s}\in\mathbb{N}^{N_{b}}$, and the unitary operator is a polaron transformation $ \hat{U}_P =\exp [\; \sum_{n} \frac{\lambda_{n} } {\Omega_{n} }\hat{S}_{n}(\hat{a}_{n}^{\dagger}-\hat{a}_{n}) \;]$. 
We want to compute expressions of the form
\begin{equation}
\hat{H}_{\boldsymbol{sr}}^{\text{cdh}}=\langle\boldsymbol{s}|\hat{U}_{P}\hat{H}\hat{U}_{P}^{\dagger}|\boldsymbol{r}\rangle,
\end{equation}
which can be written explicitly as
\begin{equation}
\begin{aligned}\hat{H}_{\boldsymbol{sr}}^{\text{cdh}} & =\langle\boldsymbol{s}|\hat{U}_{P}\hat{H}_{{\rm S}}\hat{U}_{P}^{\dagger}|\boldsymbol{r}\rangle+\sum_{n}\Big(\lambda_{n}\langle\boldsymbol{s}|\hat{U}_{P}\hat{S}_{n}(\hat{a}_{n}^{\dagger}+\hat{a}_{n})\hat{U}_{P}^{\dagger}|\boldsymbol{r}\rangle+\Omega_{n}\langle\boldsymbol{s}|\hat{U}_{P}\hat{a}_{n}^{\dagger}\hat{a}_{n}\hat{U}_{P}^{\dagger}|\boldsymbol{r}\rangle\Big).
\label{eq:AppH}
\end{aligned}
\end{equation}
We proceed to express the CDH block element $\hat{H}_{\boldsymbol{sr}}^{\text{cdh}}$ in the momentum representation. For each of the bosonic modes we have
\begin{equation}
\begin{aligned}\hat{a}_{n}^{\dagger}-\hat{a}_{n} & =-i\sqrt{\frac{2}{m_{n}\Omega_{n}}}\hat{p}_{n}=-i\sqrt{2}\hat{\tilde{p}}_{n},\\
\hat{a}_{n}^{\dagger}+\hat{a}_{n} & =\sqrt{2m_{n}\Omega_{n}}\hat{x}_{n}=\sqrt{2m_{n}\Omega_{n}}i\frac{\partial}{\partial\hat{p}_{n}}=\sqrt{2}i\frac{\partial}{\partial\hat{\tilde{p}}_{n}},\\
\hat{a}_{n}^{\dagger}\hat{a}_{n} & =\frac{1}{2}\left(\frac{\hat{p}_{n}^{2}}{m_{n}\Omega_{n}}+m_{n}\Omega_{n}\hat{x}_{n}^{2}-1\right)\\
 & =\frac{1}{2}\left(\hat{\tilde{p}}_{n}^{2}-\frac{\partial^{2}}{\partial\hat{\tilde{p}}_{n}^{2}}-1\right).
\end{aligned}
\end{equation}
Here, we define the scaled momentum as $\hat{\tilde{p}}=\hat{p}/(m\Omega)$.
The total Hamiltonian is now given by
\begin{equation}
\begin{aligned}\hat{H}_{{}} & =\hat{H}_{S}+\sum_{n}\Big(\sqrt{2}i\lambda_{n}\hat{S}_{n}\frac{\partial}{\partial\hat{\tilde{p}}_{n}}+\frac{\Omega_{n}}{2}(\hat{\tilde{p}}_{n}^{2}-\frac{\partial^{2}}{\partial\hat{\tilde{p}}_{n}^{2}}-1)\Big),
\end{aligned}
\end{equation}
and the polaron transform is
\begin{equation}
\begin{aligned}\hat{U}_P & =\exp\big(-i\sqrt{2}\sum_{n}\frac{\lambda_{n}}{\Omega_{n}}\hat{\tilde{p}}_{n}\hat{S}_{n}\big).\end{aligned}
\end{equation}
We define normalized Hermite polynomials of order $n$ to be $h_{n}(x)=(-i)^{n}H_{n}(x)/\sqrt{\sqrt{\pi}2^{n}n!}$.
(Note, $n$ here is the order of the polynomials, unrelated to $n$
used in the Hamiltonian to index modes).
We arrive at the following identities
\begin{equation}
\begin{aligned}\int h_{n}^{*}(x)h_{m}(x)e^{-x^{2}}dx & =i^{n-m}\delta_{nm},\\
\int h_{n}^{*}(x)h_{m}(x)h_{l}(x)e^{-x^{2}}dx & 
=\begin{cases}
\frac{i^{n-m-l}\sqrt{n!m!l!/\sqrt{\pi}}}{(\frac{m+n-l}{2})!(\frac{n+l-m}{2})!(\frac{m+l-n}{2})!}=i^{n-m-l}\mathcal{T}_{nml}/\sqrt[4]{\pi} & (n+m+l)|2=0 \wedge \frac{n+m+l}{2}\geq\max(n,m,l)\\
0 & \text{otherwise}
\end{cases},\\
\frac{d}{dx}h_{n}(x) & =-i\sqrt{2n}h_{n-1}(x).
\end{aligned}
\end{equation}
Here, $\mathcal{T}_{nml}$ is defined from the expression as written. Then the block elements of the extended operators $\hat{\mathcal{O}}$ are
\begin{equation}
\begin{aligned}\langle\boldsymbol{s}|\hat{U}_{P}\hat{\mathcal{O}}\hat{U}_{P}^{\dagger}|\boldsymbol{r}\rangle & =\int d\tilde{\boldsymbol{p}}d\tilde{\boldsymbol{p}}'\langle\boldsymbol{s}|\tilde{\boldsymbol{p}}\rangle\langle\tilde{\boldsymbol{p}}|\hat{U}_{P}\hat{\mathcal{O}}\hat{U}_{P}^{\dagger}|\tilde{\boldsymbol{p}}'\rangle\langle\tilde{\boldsymbol{p}}'|\boldsymbol{r}\rangle\\
 & =\int d\tilde{\boldsymbol{p}}\langle\boldsymbol{s}|\tilde{\boldsymbol{p}}\rangle\hat{U}_{P}(\tilde{\boldsymbol{p}})\hat{\mathcal{O}}(\tilde{\boldsymbol{p}})\hat{U}_{P}^{\dagger}(\tilde{\boldsymbol{p}})\langle\tilde{\boldsymbol{p}}|\boldsymbol{r}\rangle=(\hat{\mathcal{O}})_{\boldsymbol{sr}}^{\text{cdh}},
 \label{eq:AppO}
\end{aligned}
\end{equation}
where $d\tilde{\boldsymbol{p}}=\prod_{n}d\tilde{p}_{n},$ $\langle\tilde{\boldsymbol{p}}|\boldsymbol{s}\rangle=\prod_{n}h_{s_{n}}(\tilde{p}_{n})e^{-\tilde{p}_{n}^{2}/2}$ and $\hat{U}_{P}(\tilde{\boldsymbol{p}})=\langle\tilde{\boldsymbol{p}}|\hat{U}_{P}|\tilde{\boldsymbol{p}}\rangle$, and $\langle \tilde{\boldsymbol{p}}|\hat{\mathcal{O}}|{\tilde{\boldsymbol{p}}'}\rangle  = \hat{\mathcal{O}}(\tilde{\boldsymbol{p}})\delta_{
\tilde{\boldsymbol{p}}{\tilde{\boldsymbol{p}}}'}
$, since we represent operators of the modes in the momentum representation. System's operators naturally commute with the bosonic modes' operators. For simplicity, we write $\hat{U}_{P}(\tilde{\boldsymbol{p}})=\hat{U}_{P}$ from now on.

To find the CDH we have to evaluate
each term in in Eq. (\ref{eq:AppH}). 
We now evaluate three terms, labelled as $A_c$, $A_{c-s}$, and $A_{c-b}$, that build the CDH, including the leaky case (to be discussed in Sec. \ref{sec:Abaths}.
We first evaluate
\begin{equation}
\begin{aligned}
A_{c-b}\equiv
\langle\boldsymbol{s}|\hat{U}_{P}(\hat{a}_{l}^{\dagger}+\hat{a}_{l})\hat{U}_{P}^{\dagger}|\boldsymbol{r}\rangle & =\sqrt{2}i\int\prod_{n}h_{s_{n}}^{*}(\tilde{p}_{n})e^{-\tilde{p}_{n}^{2}/2}\hat{U}_{P}\frac{\partial}{\partial\tilde{p}_{l}}\Big(\hat{U}_{P}^{\dagger}\Big)\prod_{n}h_{r_{n}}(\tilde{p}_{n})e^{-\tilde{p}_{n}^{2}/2}d\tilde{\boldsymbol{p}}\\
 & +\sqrt{2}i\int\prod_{n}h_{s_{n}}^{*}(\tilde{p}_{n})e^{-\tilde{p}_{n}^{2}/2}\hat{U}_{P}\hat{U}_{P}^{\dagger}\frac{\partial}{\partial\tilde{p}_{l}}\Big(\prod_{n}h_{r_{n}}(\tilde{p}_{n})e^{-\tilde{p}_{n}^{2}/2}\Big)d\tilde{\boldsymbol{p}}\\
 & =-\frac{2\lambda_{l}}{\Omega_{l}}\int\prod_{n}h_{s_{n}}^{*}(\tilde{p}_{n})e^{-\tilde{p}_{n}^{2}/2}\hat{U}_{P}\hat{S}_{l}\hat{U}_{P}^{\dagger}\prod_{n}h_{r_{n}}(\tilde{p}_{n})e^{-\tilde{p}_{n}^{2}/2}d\tilde{\boldsymbol{p}}\\
 & +\sqrt{2}i\prod_{n\neq l}i^{s_{n}-r_{n}}\delta_{s_{n},r_{n}}\int h_{s_{l}}^{*}(\tilde{p}_{l})e^{-\tilde{p}_{l}^{2}}\left[-i\sqrt{2r_{l}}h_{r_{l}-1}(\tilde{p}_{l})-h_{r_{l}}(\tilde{p}_{l})\tilde{p}_{l}\right]d\tilde{p}_{l}\\
 & =-\frac{2\lambda_{l}}{\Omega_{l}}\int\langle\boldsymbol{s}|\tilde{\boldsymbol{p}}\rangle\hat{U}_{P}(\tilde{\boldsymbol{p}})\hat{S}_{l}\hat{U}_{P}^{\dagger}(\tilde{\boldsymbol{p}})\langle\tilde{\boldsymbol{p}}|\boldsymbol{r}\rangle d\tilde{\boldsymbol{p}}\\
 & \;+i^{s_{l}-r_{l}+1}\left[2\sqrt{r_{l}}\delta_{s_{l},r_{l}-1}-\mathcal{T}_{s_{l},r_{l},1}\right]\prod_{n\neq l}i^{s_{n}-r_{n}}\delta_{s_{n},r_{n}}\\
 & =-\frac{2\lambda_{l}}{\Omega_{l}}\left(\hat{S}_{l}\right)_{\boldsymbol{sr}}^{\text{cdh}}+\sqrt{\text{max}(s_{l},r_{l})}\delta_{|s_{l}-r_{l}|,1}\prod_{n\neq l}\delta_{s_{n},r_{n}}\\
 & =-\frac{2\lambda_{l}}{\Omega_{l}}\left(\hat{S}_{l}\right)_{\boldsymbol{sr}}^{\text{cdh}}+\hat{D}_{s_{l}r_{l}}\prod_{n\neq l}\delta_{s_{n},r_{n}}
\end{aligned}
\end{equation}
%
%
Here, $\hat{D}_{s_{l}r_{l}}$ is introduced as a short notation. 
In the derivation, we used the fact that 
\begin{equation}
\begin{aligned}\frac{\partial}{\partial\tilde{p}_{l}}\Big(\prod_{n}h_{r_{n}}(\tilde{p}_{n})e^{-\tilde{p}_{n}^{2}/2}\Big) & =\left[\frac{\partial}{\partial\tilde{p}_{l}}\left(h_{r_{l}}(\tilde{p}_{l})\right)e^{-\tilde{p}_{l}^{2}/2}+h_{r_{l}}(\tilde{p}_{l})\frac{\partial}{\partial\tilde{p}_{l}}\left(e^{-\tilde{p}_{l}^{2}/2}\right)\right]\prod_{n\neq l}h_{r_{n}}(\tilde{p}_{n})e^{-\tilde{p}_{n}^{2}/2}\\
 & =\left[-i\sqrt{2r_{l}}h_{r_{l}-1}(\tilde{p}_{l})e^{-\tilde{p}_{l}^{2}/2}-\tilde{p}_{l}h_{r_{l}}(\tilde{p}_{l})e^{-\tilde{p}_{l}^{2}/2}\right]\prod_{n\neq l}h_{r_{n}}(\tilde{p}_{n})e^{-\tilde{p}_{n}^{2}/2}.
\end{aligned}
\end{equation}
The second term we calculate involve the term coupling the bosonic mode and the system,
\begin{equation}
\begin{aligned}
A_{c-s}\equiv
\langle\boldsymbol{s}|\hat{U}_{P}\hat{S}_{l}(\hat{a}_{l}^{\dagger}+\hat{a}_{l})\hat{U}_{P}^{\dagger}|\boldsymbol{r}\rangle & =\sqrt{2}i\int\prod_{n}h_{s_{n}}^{*}(\tilde{p}_{n})e^{-\tilde{p}_{n}^{2}/2}\hat{U}_{P}\hat{S}_{l}\frac{\partial}{\partial\tilde{p}_{l}}\Big(\hat{U}_{P}^{\dagger}\Big)\prod_{n}h_{r_{n}}(\tilde{p}_{n})e^{-\tilde{p}_{n}^{2}/2}d\tilde{\boldsymbol{p}}=\\
 & +\sqrt{2}i\int\prod_{n}h_{s_{n}}^{*}(\tilde{p}_{n})e^{-\tilde{p}_{n}^{2}/2}\hat{U}_{P}\hat{S}_{l}\hat{U}_{P}^{\dagger}\frac{\partial}{\partial\tilde{p}_{l}}\Big(\prod_{n}h_{r_{n}}(\tilde{p}_{n})e^{-\tilde{p}_{n}^{2}/2}\Big)d\tilde{\boldsymbol{p}}\\
 & =-\frac{2\lambda_{l}}{\Omega_{l}}\left(\hat{S}_{l}^{2}\right)_{\boldsymbol{sr}}^{\text{cdh}}\\
 & +\sqrt{2}i\int\prod_{n}h_{s_{n}}^{*}(\tilde{p}_{n})e^{-\tilde{p}_{n}^{2}/2}\hat{U}_{P}\hat{S}_{l}\hat{U}_{P}^{\dagger}\frac{\partial}{\partial\tilde{p}_{l}}\Big(\prod_{n}h_{r_{n}}(\tilde{p}_{n})e^{-\tilde{p}_{n}^{2}/2}\Big)d\tilde{\boldsymbol{p}}.
\end{aligned}
\end{equation}
Lastly, we compute the term containing the number operator,
\begin{equation}
\begin{aligned}
A_{c}\equiv
\langle\boldsymbol{s}|\hat{U}_{P}\hat{a}_{l}^{\dagger}\hat{a}_{l}\hat{U}_{P}^{\dagger}|\boldsymbol{r}\rangle & =\int\prod_{n}h_{s_{n}}^{*}(\tilde{p}_{n})e^{-\tilde{p}_{n}^{2}/2}\hat{U}_{P}(\tilde{\boldsymbol{p}})\frac{1}{2}(\tilde{p_{l}}^{2}-\frac{\partial^{2}}{\partial\tilde{p}_{l}^{2}}-1)\Big(\hat{U}_{P}^{\dagger}(\tilde{\boldsymbol{p}})\prod_{n}h_{r_{n}}(\tilde{p}_{n})e^{-\tilde{p}_{n}^{2}/2}\Big)d\tilde{\boldsymbol{p}}\\
 & =\frac{1}{2}\int\prod_{n}h_{s_{n}}^{*}(\tilde{p}_{n})e^{-\tilde{p}_{n}^{2}/2}\left(\tilde{p_{l}}^{2}-1\right)\prod_{n}h_{r_{n}}(\tilde{p}_{n})e^{-\tilde{p}_{n}^{2}/2}d\tilde{\boldsymbol{p}}\\
 & -\frac{1}{2}\int\prod_{n}h_{s_{n}}^{*}(\tilde{p}_{n})e^{-\tilde{p}_{n}^{2}/2}\hat{U}_{P}(\tilde{\boldsymbol{p}})\frac{\partial^{2}}{\partial\tilde{p}_{l}^{2}}\Big(\hat{U}_{P}^{\dagger}(\tilde{\boldsymbol{p}})\Big)\prod_{n}h_{r_{n}}(\tilde{p}_{n})e^{-\tilde{p}_{n}^{2}/2}d\tilde{\boldsymbol{p}}\\
 & -\frac{1}{2}\int\prod_{n}h_{s_{n}}^{*}(\tilde{p}_{n})e^{-\tilde{p}_{n}^{2}/2}\frac{\partial^{2}}{\partial\tilde{p}_{l}^{2}}\Big(\prod_{n}h_{r_{n}}(\tilde{p}_{n})e^{-\tilde{p}_{n}^{2}/2}\Big)d\tilde{\boldsymbol{p}}\\
 & -\int\prod_{n}h_{s_{n}}^{*}(\tilde{p}_{n})e^{-\tilde{p}_{n}^{2}/2}\hat{U}_{P}(\tilde{\boldsymbol{p}})\frac{\partial}{\partial\tilde{p}_{l}}\Big(\hat{U}_{P}^{\dagger}(\tilde{\boldsymbol{p}})\Big)\frac{\partial}{\partial\tilde{p}_{l}}\Big(\prod_{n}h_{r_{n}}(\tilde{p}_{n})e^{-\tilde{p}_{n}^{2}/2}\Big)d\tilde{\boldsymbol{p}}\\
 & =\frac{\lambda_{l}^{2}}{\Omega_{l}^{2}}\int\langle\boldsymbol{s}|\tilde{\boldsymbol{p}}\rangle\hat{U}_{P}(\tilde{\boldsymbol{p}})\hat{S}_{l}^{2}\hat{U}_{P}^{\dagger}(\tilde{\boldsymbol{p}})\langle\tilde{\boldsymbol{p}}|\boldsymbol{r}\rangle d\tilde{\boldsymbol{p}}\\
 & -\frac{1}{2}\prod_{n\neq l}i^{s_{n}-r_{n}}\delta_{s_{n},r_{n}}\int h_{s_{l}}^{*}(\tilde{p}_{l})e^{-\tilde{p}_{l}^{2}}\left[-\sqrt{4r_{l}(r_{l}-1)}h_{r_{l}-2}(\tilde{p}_{l})+i\sqrt{2r_{l}}2\tilde{p}_{l}h_{r_{l}-1}(\tilde{p}_{l})\right]d\tilde{p}_{l}\\
 & -i\sqrt{2}\int\prod_{n}h_{s_{n}}^{*}(\tilde{p}_{n})e^{-\tilde{p}_{n}^{2}/2}\hat{U}_{P}(\tilde{\boldsymbol{p}})\hat{S}_{l}\hat{U}_{P}^{\dagger}(\tilde{\boldsymbol{p}})\frac{\partial}{\partial\tilde{p}_{l}}\Big(\prod_{n}h_{r_{n}}(\tilde{p}_{n})e^{-\tilde{p}_{n}^{2}/2}\Big)d\tilde{\boldsymbol{p}}\\
 & =\frac{\lambda_{l}^{2}}{\Omega_{l}^{2}}\left(\hat{S}_{l}^{2}\right)_{\boldsymbol{s,r}}^{\text{cdh}}+\prod_{n\neq l}\delta_{s_{n},r_{n}}i^{s_{l}-r_{l}}\sqrt{r_{l}}\left[\mathcal{T}_{s_{l},r_{l}-1,1}-\sqrt{(r_{l}-1)}\delta_{s_{l},r_{l}-2}\right]\\
 & -i\sqrt{2}\frac{\lambda_l}{\Omega_l}\int\prod_{n}h_{s_{n}}^{*}(\tilde{p}_{n})e^{-\tilde{p}_{n}^{2}/2}\hat{U}_{P}(\tilde{\boldsymbol{p}})\hat{S}_{l}{U}_{P}^{\dagger}(\tilde{\boldsymbol{p}})\frac{\partial}{\partial\tilde{p}_{l}}\Big(\prod_{n}h_{r_{n}}(\tilde{p}_{n})e^{-\tilde{p}_{n}^{2}/2}\Big)d\tilde{\boldsymbol{p}}\\
 & =\frac{\lambda_{l}^{2}}{\Omega_{l}^{2}}\left(\hat{S}_{l}^{2}\right)_{\boldsymbol{s,r}}^{\text{cdh}}+\text{max(}s_{l},r_{l})\delta_{\boldsymbol{s},\boldsymbol{r}}\\
 & -i\sqrt{2}\frac{\lambda_l}{\Omega_l}\int\prod_{n}h_{s_{n}}^{*}(\tilde{p}_{n})e^{-\tilde{p}_{n}^{2}/2}\hat{U}_{P}(\tilde{\boldsymbol{p}})\hat{S}_{l}\hat{U}_{P}^{\dagger}(\tilde{\boldsymbol{p}})\frac{\partial}{\partial\tilde{p}_{l}}\Big(\prod_{n}h_{r_{n}}(\tilde{p}_{n})e^{-\tilde{p}_{n}^{2}/2}\Big)d\tilde{\boldsymbol{p}},
\end{aligned}
\end{equation}
where we cancelled the two terms proportional to $\tilde{p}_{l}^{2}-1$
and used
\begin{equation}
\begin{aligned}\frac{\partial^{2}}{\partial\tilde{p}_{l}^{2}}\Big(h_{r_{l}}(\tilde{p}_{l})e^{-\tilde{p}_{l}^{2}/2}\Big) & =\frac{\partial^{2}}{\partial\tilde{p}_{l}^{2}}\left(h_{r_{l}}(\tilde{p}_{l})+e^{-\tilde{p}_{l}^{2}/2}\right)+2\frac{\partial}{\partial\tilde{p}_{l}}\left(h_{r_{l}}(\tilde{p}_{l})\right)\frac{\partial}{\partial\tilde{p}_{l}}\left(e^{-\tilde{p}_{l}^{2}/2}\right)\\
 & =\frac{\partial}{\partial\tilde{p}_{l}}\left[-i\sqrt{2r_{l}}h_{r_{l}-1}(\tilde{p}_{l})e^{-\tilde{p}_{l}^{2}/2}-\tilde{p}_{l}h_{r_{l}}(\tilde{p}_{l})e^{-\tilde{p}_{l}^{2}/2}\right]\\
 & =-\sqrt{4r_{l}(r_{l}-1)}h_{r_{l}-2}(\tilde{p}_{l})e^{-\tilde{p}_{l}^{2}/2}+i2\sqrt{2r_{l}}\tilde{p}_{l}h_{r_{l}-1}(\tilde{p}_{l})e^{-\tilde{p}_{l}^{2}/2}\\
 & \;+(\tilde{p}_{l}^{2}-1)h_{r_{l}}(\tilde{p}_{l})e^{-\tilde{p}_{l}^{2}/2}.
\end{aligned}
\end{equation}
We now construct the CDH (\ref{eq:AppH}), based on combining $A_{c}$ and $A_{c-s}$, along with the appropriate prefactors. We find that the CDH block element is
\begin{equation}
\begin{aligned}\hat{H}_{\boldsymbol{sr}}^{\text{cdh}} & =\left(\hat{H}_{{\rm S}}-\sum_{n}\left[\frac{\lambda_{n}^{2}}{\Omega_{n}}\hat{S}_{n}^{2}-\Omega_{n}s_{n}\delta_{\boldsymbol{s},\boldsymbol{r}}\right]\right)_{\boldsymbol{sr}}^{\text{cdh}}.
\label{eq:AppHcdh}
\end{aligned}
\end{equation}
We introduce a notation for the excitation number operator, $ (s_n\delta_{\boldsymbol{s},\boldsymbol{r}})^{\text{cdh}}_{\boldsymbol{s},\boldsymbol{r}} = \hat{N}_n$, and for a mapped system operator $\hat{\mathcal{O}}$ obtained after the application of the polaron transform and truncation to the lowest $M$ Fock states 
\begin{equation}
\begin{aligned}\left(\hat{\mathcal{O}}\right)_{(M)}^{\text{cdh}}=\sum_{\boldsymbol{s},\boldsymbol{r}}|\boldsymbol{s}\rangle\langle\boldsymbol{r}|\otimes\left(\hat{\mathcal{O}}\right)_{\boldsymbol{s}\boldsymbol{r}}^{\text{cdh}},\end{aligned}
\end{equation}
where $\forall_{n}\;s_{n},r_{n}\leq M$. 

Equation (\ref{eq:AppHcdh}) is a central analytical result of this work, providing the form of the mapped-truncated Hamiltonian terms.
For concrete models, one needs to evaluate system's operators in the CDH representation. This can be achieved in two ways, the Integration method and the Spectral decomposition method.

\subsubsection{Integration method}
CDH operators of the system can be obtained by evaluating, either analytically or numerically, the integrals in momentum representation.
Using
$\hat{\mathcal{O}}$ to represent an operator of the system, we evaluate it from
\begin{equation}
\begin{aligned}\langle\boldsymbol{s}|\hat{U}_{P}\hat{\mathcal{O}}\hat{U}_{P}^{\dagger}|\boldsymbol{r}\rangle &  & =\int d\tilde{\boldsymbol{p}}\langle\boldsymbol{s}|\tilde{\boldsymbol{p}}\rangle\hat{U}_{P}(\tilde{\boldsymbol{p}})\hat{\mathcal{O}}\hat{U}_{P}^{\dagger}(\tilde{\boldsymbol{p}})\langle\tilde{\boldsymbol{p}}|\boldsymbol{r}\rangle.
 \label{eq:AppO2}
\end{aligned}
\end{equation}
See concrete expressions for the integrand below Eq. (\ref{eq:AppO}).

\subsubsection{Spectral Decomposition Method}

Alternatively, if the system's coupling operators commute, one can use Eq.(3) from the main text to compute CDH blocks of an operator $\hat{\mathcal{O}}$. First, we note that in this case we can factor the mixed polaron transform into consecutive, single mode shifts. Since each of the factors can be applied consecutively, we focus on the action of a single mode transform. We write
\begin{equation}
    \begin{aligned}
        \langle s|\hat{U}_{P}\hat{\mathcal{O}}\hat{U}_{P}^{\dagger}|r\rangle&=\langle s|\left(\sum_{k}|k\rangle\langle k|\otimes\hat{D}(\epsilon s_{k})\right)\hat{\mathcal{O}}\left(\sum_{k'}|k'\rangle\langle k'|\otimes\hat{D}(-\epsilon s_{k'})\right)|r\rangle\\&=\sum_{k,k'}\left(|k\rangle\langle k|\hat{\mathcal{O}}|k'\rangle\langle k'|\right)\otimes\left(\langle s|\hat{D}\left(\epsilon(s_{k}-s_{k'})\right)|r\rangle\right),\label{eq:rotation_with_displacement}
    \end{aligned}
\end{equation}
where matrix elements of the displacement operator are
\begin{equation}
    \begin{aligned}
        \langle s|\hat{D}(\alpha)|r\rangle&=\begin{cases}
\sqrt{\frac{r!}{s!}}\alpha^{s-r}L_{r}^{(s-r)}(|\alpha|^{2})e^{-|\alpha|^{2}/2} & s\geq r\\
\sqrt{\frac{s!}{r!}}(-\alpha^{*})^{r-s}L_{s}^{(r-s)}(|\alpha|^{2})e^{-|\alpha|^{2}/2} & s<r.
\end{cases}
    \end{aligned}
\end{equation}
Depending on the convenience of calculations, one can use this method, or compute momentum integrals in the Integration method.

\subsection{Adding a dissipative bath}
\label{sec:Abaths}

We now consider an extension to systems, where the cavity modes are additionally each coupled to their own bosonic bath leading to dissipation,
\begin{equation}
\begin{aligned}\hat{H}_{{}} & =\hat{H}_{S}+\sum_{n}\Big(\lambda_{n}\hat{S}_{n}(\hat{a}_{n}^{\dagger}+\hat{a}_{n})+\Omega_{n}\hat{a}_{n}^{\dagger}\hat{a}_{n}\Big)\\
 & +\sum_{n}{\sum_{k}}\Big(f_{n,k}(\hat{a}_{n}^{\dagger}+\hat{a}_{n})(\hat{b}_{n,k}^{\dagger}+\hat{b}_{n,k})+\omega_{n,k}\hat{b}_{n,k}^{\dagger}\hat{b}_{n,k}\Big).
\end{aligned}
\end{equation}
Here, $f_{n,k}$, $\omega_{n,k}$, and $\hat{b}_{n,k}$ are coupling strength, frequencies and annihilation operators of the dissipative bath modes, respectively. Note that each cavity mode ($\hat a_n$) is coupled to an independent collections of boson modes ($\hat b_{n,k}$).
The new terms can be interpreted as introducing leakage to the cavity, or as extending the model to that of a system originally coupled to a thermal bath, after a reaction coordinate transformation \cite{Nazir2018,anto2023effective}. The CDH mapping of this system-leaky cavity Hamiltonian is
\begin{equation}
\begin{aligned}\hat{H}_{\boldsymbol{sr}}^{\text{cdh}} & =\langle\boldsymbol{s}|\hat{U}_{P}\hat{H}_{{\rm S}}\hat{U}_{P}^{\dagger}|\boldsymbol{r}\rangle+\sum_{n}\Big(\lambda_{n}\langle\boldsymbol{s}|\hat{U}_{P}\hat{S}_{n}(\hat{a}_{n}^{\dagger}+\hat{a}_{n})\hat{U}_{P}^{\dagger}|\boldsymbol{r}\rangle+\Omega_{n}\langle\boldsymbol{s}|\hat{U}_{P}\hat{a}_{n}^{\dagger}\hat{a}_{n}\hat{U}_{P}^{\dagger}|\boldsymbol{r}\rangle\Big)\\
 & +\sum_{n}{\sum_{k}}\Big(f_{n,k}\langle\boldsymbol{s}|\hat{U}_{P}(\hat{a}_{n}^{\dagger}+\hat{a}_{n})\hat{U}_{P}^{\dagger}|\boldsymbol{r}\rangle(\hat{b}_{n,k}^{\dagger}+\hat{b}_{n,k})+\omega_{n,k}\hat{b}_{n,k}^{\dagger}\hat{b}_{n,k}\Big).
\end{aligned}
\end{equation}
Compared to the case without leakage, we recognize one additional term, which we map to the momentum representation and obtain,
\begin{equation}
\begin{aligned}\hat{H}_{{}} & =\hat{H}_{S}+\sum_{n}\Big(\sqrt{2}i\lambda_{n}\hat{S}_{n}\frac{\partial}{\partial\hat{\tilde{p}}_{n}}+\frac{\Omega_{n}}{2}(\hat{\tilde{p}}_{n}^{2}-\frac{\partial^{2}}{\partial\hat{\tilde{p}}_{n}^{2}}-1)\Big)\\
 & +\sum_{n}{\sum_{k}}\Big(\sqrt{2}if_{n,k}\frac{\partial}{\partial\hat{\tilde{p}}_{n}}(\hat{b}_{n,k}^{\dagger}+\hat{b}_{n,k})+\omega_{n,k}\hat{b}_{n,k}^{\dagger}\hat{b}_{n,k}\Big).
\end{aligned}
\end{equation}
Using our expressions, $A_{c-b}$, $A_{c}$ and $A_{c-s}$, we arrive at
\begin{equation}
\begin{aligned}\hat{H}_{\boldsymbol{sr}}^{\text{cdh}} & =\left(\hat{H}_{{\rm S}}-\sum_{n}\left[\frac{\lambda_{n}^{2}}{\Omega_{n}}\hat{S}_{n}^{2}-\Omega_{n}s_{n}\delta_{\boldsymbol{s},\boldsymbol{r}}\right]\right.\\
 & \left.+\sum_{n}{\sum_{k}}\left[\omega_{n,k}\hat{b}_{n,k}^{\dagger}\hat{b}_{n,k}-f_{n,k}\left[\frac{2\lambda_{n}}{\Omega_{n}}\hat{S}_{n}-\hat{D}_{s_{n}r_{n}}\right](\hat{b}_{n,k}^{\dagger}+\hat{b}_{n,k})\right]\right)_{\boldsymbol{sr}}^{\text{cdh}}
\end{aligned}
\end{equation}
In sum, the mapping in the case with leakage is equivalent to making the following substitutions in the original Hamiltonian
\begin{equation}
\begin{aligned}\hat{H}_{{\rm S}} & \rightarrow\left(\hat{H}_{{\rm S}}\right)_{\boldsymbol{sr}}^{\text{cdh}}-\sum_{n}\left[\frac{\lambda_{n}^{2}}{\Omega_{n}}\left(\hat{S}_{n}^{2}\right)_{\boldsymbol{sr}}^{\text{cdh}}-\Omega_{n}\hat{N}_n\right],\\
f_{n,k}(\hat{a}_{n}^{\dagger}+\hat{a}_{n}) & \rightarrow-f_{n,k}\left[\frac{2\lambda_{n}}{\Omega_{n}}\left(\hat{S}_{n}\right)_{\boldsymbol{sr}}^{\text{cdh}}-\hat{D}_{s_{n}r_{n}}\right].
\end{aligned}
\end{equation}
%

\subsection{Numerical Simulations}
\label{sec:Anum}
The entangling transformation is unitary and therefore preserves the eigenvalues of the Hamiltonian. Its advantage lies in accelerating convergence: when the transformed Hamiltonian is truncated, yielding what we refer to as the CDH representation, it provides a significantly more accurate approximation to the spectrum than an equivalently truncated untransformed Hamiltonian, as demonstrated in Fig. 2 (main).

We further calculate expectation values with respect to the equilibrium state, with $\beta$ as the inverse temperature. We exemplify the process on the magnetization in the $z$ orientation,
\bea
\langle \hat \sigma_z\rangle_{eq}
&=&
\frac{1}{Z}{\rm Tr}[e^{-\beta \hat H} \hat \sigma_z]
\nonumber\\
&=&\frac{1}{Z}{\rm Tr}[\hat U_P e^{-\beta \hat H} \hat U_P^{\dagger} \hat U_P\hat \sigma_z \hat U_P^{\dagger}]
\nonumber\\
&=&\frac{1}{Z}{\rm Tr}[ e^{-\beta \hat U_P\hat H\hat U_P^{\dagger}}  \hat U_P\hat \sigma_z \hat U_P^{\dagger}].
\eea
Here, $Z={\rm Tr}[e^{-\beta \hat H}]$ is the partition function.
In simulations,  we represent 
$\hat U_P\hat H\hat U_P^{\dagger}$
by $M$ levels for the cavity, which is the CDH Hamiltonian, 
thus,
$\rho_P = \frac{1}{Z} e^{-\beta \hat H_{(M)}^{{\rm cdh}}}$
with the partition function calculated for the CDH.

To calculate the expectation values, we need to perform a transformation of the operator $\hat U_P\sigma_z \hat U_P^{\dagger}$. Although intuitively, this transformation would be performed using $M$ levels for the cavity to represent the polaron operators, similarly to the dimension of the CDH, we found that in ``difficult" parameter ranges (low temperature and resonant conditions) one needed to perform this transformation at a higher dimension, $M_P>M$. The resulting matrix, after the transformation, is truncated to include $M$ modes,  matching the dimension of the density matrix, $\rho_P$. We identify  ``difficult" parameter ranges with two observations: Convergence with increasing $M$ is not monotonic, and the deep-strong coupling  limit (which is often easy to physically understand) is incorrect/nonphysical. In those cases, we extend $M_P$ until we recover both monotonic convergence with $M$, and a proper physical behavior when $\lambda\to \infty$.

The CDH Hamiltonian is primarily suited to capture energetics and matter dressing by the cavity, since these effects are dominated by low cavity occupations, even in the ultrastrong and deep-strong coupling. This is because the polaron transformation absorbs the large displacement exactly. Rotated observables, on the other hand, probe fluctuations and correlations in the original (lab) frame. These generally involve higher-order processes in the bosonic displacement operator, which are more sensitive to the tails of the cavity Fock space. For observables commuting with the polaron unitary, this problem does not apply, since the unitary can be removed from the trace by the cyclic property of the trace.

The CDH mapping can be done purely numerically, rather than analytically. To do so, one needs to construct the mixed polaron unitary with $M_P$ lowest levels, rotate the system Hamiltonian, add the terms ``$\hat{S}^2$" and ``$\hat{N}$" and finally truncate to the desired order $M$. The most expensive operation in this procedure is matrix multiplication, which scales like $O(M^3)$. There are no restrictions on when the method can be applied purely numerically. Here, we choose to present the analytical expressions for system operators in different cavity sectors in order to gain intuition on physical processes in the system, mediated by cavity photons.

Returning to the construction of the unitary polaron, in the case of commuting coupling operators, the computation is straightforward. The only nontrivial part is to  diagonalize the coupling operators (also $O(M^3)$ complexity) as in Eq. (27) and Eq. (28).

When system coupling operators do not commute, one has to evaluate integrals over momenta as in Eq. (26). For each momentum integration, the integrand is of the following form: Gaussian envelope $\times$ product of two Hermite polynomials $\times$ rotated system operator in the momentum basis, $\hat{U}_P(p) \hat{\mathcal{O}} \hat{U}_P^\dagger(p)$. The integrand is a system operator, and it can be evaluated element by element in its matrix representation, so the complexity is $O(M^2)$, times the complexity of the integration, which we observe to scale slower than the rest of the procedure.

These momentum integrals can be difficult to find analytically, but for the spin one half and spin one, analytical expressions can be found using the Euler formula or using results from \cite{CURTRIGHT2015401} for the spin one case. These results do not depend on the number of spins, but rather on the spin value.

\section{Quantum Rabi model}
\label{sec:AppB}

\subsection{Hamiltonian mapping}
We use here the general theory developed in Sec. \ref{sec:AppA} to obtain the CDH of the quantum Rabi model. We need to evaluate the following integrals,
\begin{equation}
\left(\hat{\sigma}^{z}\right)_{sr}^{\text{cdh}}=\int_{-\infty}^{\infty}\hat{U}_{P}\hat{\sigma}^{z}\hat{U}_{P}^{\dagger}h_{s}^{*}(p)h_{r}(p)e^{-p^{2}}dp,
\end{equation}
where $H_{n}(p)$ are Hermite polynomials, $\epsilon=\lambda/\Omega$, and $\hat{U}_{P}=\exp(-i\sqrt{2}\epsilon p\hat{\sigma}^{x})$. Alternatively, using Eq.(\ref{eq:rotation_with_displacement}) we write
\begin{equation}
    \begin{aligned}
       \left(\hat{\sigma}^{z}\right)_{sr}^{\text{cdh}}&=\sum_{k\neq k'}\left(-2e^{-2\epsilon^{2}}|k\rangle\langle k'|\right)\otimes\left(\begin{cases}
\sqrt{\frac{s!}{r!}}\left(\epsilon(s_{k}-s_{k'})\right)^{r-s}L_{s}^{(r-s)}(4\epsilon^{2}) & s\geq r\\
\sqrt{\frac{r!}{s!}}\left(\epsilon(s_{k'}-s_{k})\right)^{s-r}L_{r}^{(s-r)}(4\epsilon^{2}) & s<r
\end{cases}\right),
    \end{aligned}
\end{equation}
where $s_k,|k\rangle$ are eigenvalues and eigenvectors of $\hat{\sigma}^x$. We now construct the operator by truncating the mode to four levels, letting $0 \leq s,r\leq 3$. Note that in order to relate results to the matrix elements, we enumerate cavity levels from $0$ to $M-1$, with, e.g., $\left(\hat{\sigma}^{z}\right)_{(1)}^{\text{cdh}}$   corresponding to maintaining only the ground level of the cavity, and $\left(\hat{\sigma}^{z}\right)_{(4)}^{\text{cdh}}$ corresponding to maintaining 4 levels, starting from the ground state. We readily obtain 
\begin{equation}
\begin{aligned}\left(\hat{\sigma}^{z}\right)_{(4)}^{\text{cdh}} & =\sum_{s,r=0}^{3}|s\rangle\langle r|\otimes\left(\hat{\sigma}^{z}\right)_{sr}^{\text{cdh}}\\
 & =e^{-2\epsilon^{2}}\begin{pmatrix}\hat{\sigma}^{z} & 2\epsilon i\hat{\sigma}^{y} & 2\sqrt{2}\epsilon^{2}\hat{\sigma}^{z} & 4\sqrt{\frac{2}{3}}\epsilon^{3}i\hat{\sigma}^{y}\\
-2\epsilon i\hat{\sigma}^{y} & \left(1-4\epsilon^{2}\right)\hat{\sigma}^{z} & -2\sqrt{2}\epsilon\left(2\epsilon^{2}-1\right)i\hat{\sigma}^{y} & -2\sqrt{\frac{2}{3}}\epsilon^{2}\left(4\epsilon^{2}-3\right)\hat{\sigma}^{z}\\
2\sqrt{2}\epsilon^{2}\hat{\sigma}^{z} & 2\sqrt{2}\epsilon\left(2\epsilon^{2}-1\right)i\hat{\sigma}^{y} & \left(8\epsilon^{4}-8\epsilon^{2}+1\right)\hat{\sigma}^{z} & \frac{2\epsilon\left(8\epsilon^{4}-12\epsilon^{2}+3\right)}{\sqrt{3}}i\hat{\sigma}^{y}\\
-4\sqrt{\frac{2}{3}}\epsilon^{3}i\hat{\sigma}^{y} & -2\sqrt{\frac{2}{3}}\epsilon^{2}\left(4\epsilon^{2}-3\right)\hat{\sigma}^{z} & -\frac{2\epsilon\left(8\epsilon^{4}-12\epsilon^{2}+3\right)}{\sqrt{3}}i\hat{\sigma}^{y} & \frac{1}{3}\left(-32\epsilon^{6}+72\epsilon^{4}-36\epsilon^{2}+3\right)\hat{\sigma}^{z}
\end{pmatrix}.
\end{aligned}
\end{equation}
Putting together all terms, the fully mapped Rabi Hamiltonian is given by
{\tiny
\begin{equation}
    \begin{aligned}
        \left(\hat{H}\right)_{(4)}^{\text{cdh}}=\begin{pmatrix}\tilde{\Delta}\hat{\sigma}^{z}-\frac{\lambda^{2}}{\Omega} & \tilde{\Delta}2\epsilon i\hat{\sigma}^{y} & \tilde{\Delta}2\sqrt{2}\epsilon^{2}\hat{\sigma}^{z} & \tilde{\Delta}4\sqrt{\frac{2}{3}}\epsilon^{3}i\hat{\sigma}^{y}\\
-\tilde{\Delta}2\epsilon i\hat{\sigma}^{y} & \tilde{\Delta}\left(1-4\epsilon^{2}\right)\hat{\sigma}^{z}+\Omega-\frac{\lambda^{2}}{\Omega} & -\tilde{\Delta}2\sqrt{2}\epsilon\left(2\epsilon^{2}-1\right)i\hat{\sigma}^{y} & -\tilde{\Delta}2\sqrt{\frac{2}{3}}\epsilon^{2}\left(4\epsilon^{2}-3\right)\hat{\sigma}^{z}\\
\tilde{\Delta}2\sqrt{2}\epsilon^{2}\hat{\sigma}^{z} & \tilde{\Delta}2\sqrt{2}\epsilon\left(2\epsilon^{2}-1\right)i\hat{\sigma}^{y} & \tilde{\Delta}\left(8\epsilon^{4}-8\epsilon^{2}+1\right)\hat{\sigma}^{z}+2\Omega-\frac{\lambda^{2}}{\Omega} & \tilde{\Delta}\frac{2\epsilon\left(8\epsilon^{4}-12\epsilon^{2}+3\right)}{\sqrt{3}}i\hat{\sigma}^{y}\\
-\tilde{\Delta}4\sqrt{\frac{2}{3}}\epsilon^{3}i\hat{\sigma}^{y} & -\tilde{\Delta}2\sqrt{\frac{2}{3}}\epsilon^{2}\left(4\epsilon^{2}-3\right)\hat{\sigma}^{z} & -\tilde{\Delta}\frac{2\epsilon\left(8\epsilon^{4}-12\epsilon^{2}+3\right)}{\sqrt{3}}i\hat{\sigma}^{y} & \tilde{\Delta}\frac{1}{3}\left(-32\epsilon^{6}+72\epsilon^{4}-36\epsilon^{2}+3\right)\hat{\sigma}^{z}+3\Omega-\frac{\lambda^{2}}{\Omega}
\end{pmatrix}.
    \end{aligned}
\end{equation}
}
Here, $\tilde{\Delta} = e^{-2\epsilon^2} \Delta$. These expressions are used to compute the CDH spectra and observables in Fig.1 in the main text. We note that in Ref. \citenum{PhysRevA.108.043717}
a similar method was used to find the spectrum of the quantum Rabi model. The key difference was the approximation used there, which consisted of considering only the diagonal and first off-diagonal terms in the polaron transform in the number basis. Here, the approximation is based on discarding terms corresponding to higher excited states in the polaron basis. Both methods provide excellent
convergence. However, our approach does not rely on the form of the polaron exclusive to the particular system coupling operator used, thus it can be readily applied to other light-matter models beyond the Rabi model.


\subsection{Matter properties: Ground state and thermal magnetization}

\begin{figure*}[hbpt]
\centering \includegraphics[width=0.9\textwidth]{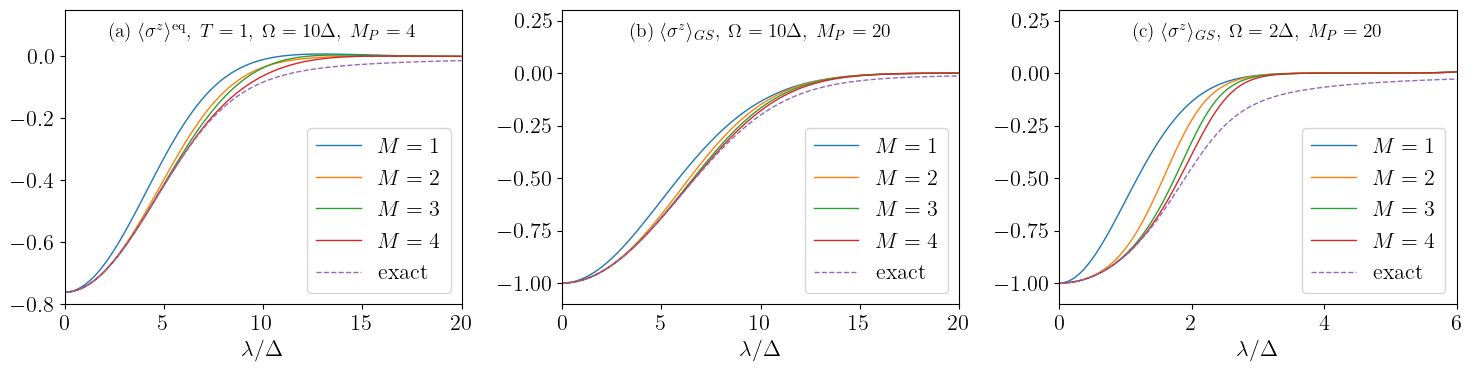} \caption{Average magnetization, $\langle \hat{\sigma}^z \rangle$, as a function of the normalized coupling strength, $\lambda / \Omega$ (a) at equilibrium, $T=1$, and off resonance $\Omega=10\Delta$, (b) at zero temperature and off resonance $\Omega=10\Delta$, and (c) at
zero temperature, in resonance, $\Omega=2\Delta$.
Colors indicate CDH results for increasing $M$ values. Converged numerical results from the bare representation appear in dashed. We used $\Delta=1$.
}
\label{fig:various_states} 
\end{figure*}

\begin{figure*}[htpb]
\centering \includegraphics[width=0.9\textwidth]{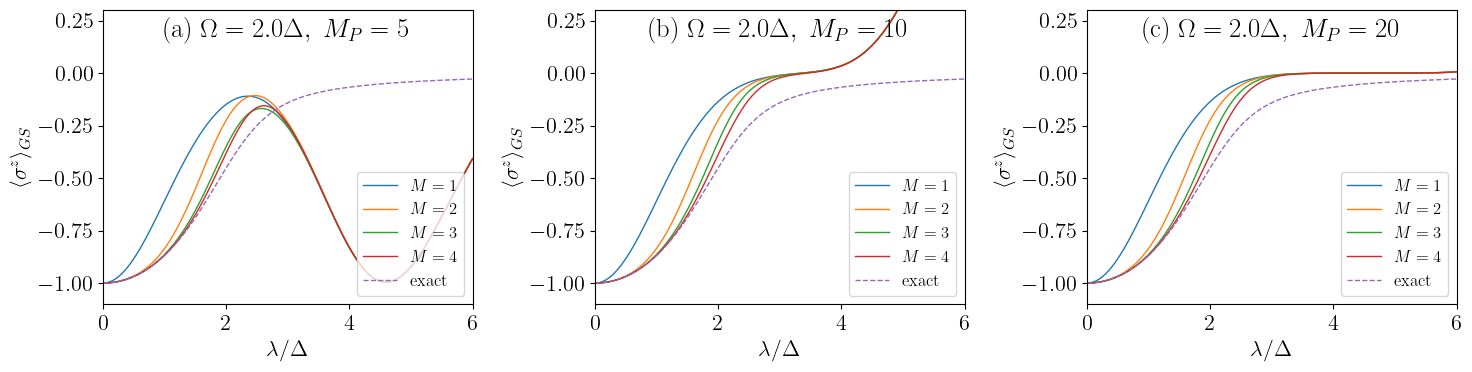} \caption{Average ground state magnetization, $\langle \hat{\sigma}^z \rangle_{GS}$, as a function of the normalized coupling strength, $\lambda / \Omega$, at resonance $\Omega = 2\Delta$.
We vary  $M_{P}$, the dimensionality of the cavity Hilbert space used to rotate operators with (a) $M_P=5$, (b) $10$, and (c) $20$. Colors indicate CDH results for different $M$. Converged numerical results appear in dashed. We used $\Delta=1$.
}
\label{fig:polaron_truncation} 
\end{figure*}

In Fig.~\ref{fig:various_states}, we plot the average magnetization as a function of the normalized coupling strength, $\lambda / \Omega$, for different scenarios: (a) finite temperature and off resonance, $\Delta \ll \Omega$; (b) zero temperature and off resonance, $\Delta \ll \Omega$; and (c) zero temperature at resonance. Comparing the results in panels (a)–(b) with panel (c) and Fig.1 in the main text, we observe that the CDH representation achieves faster convergence when moving away from resonance in parameter space. We observe slower convergence with $M$ for the ground state observables in both resonance and off resonance cases: For thermal states, $M_{P}=M$ already yields optimal convergence. For ground state calculations, $M_{P}=20$ is needed, which equals the dimension that makes the bare representation calculations converge on the shown range of $\lambda$, while $M=4$.
As a reminder, see Sec. \ref{sec:Anum}, $M$ corresponds to the number of cavity states maintained in the construction of the CDH. $M_P$ describes the number of cavity states used to prepare operators in the 
transformed basis  (later truncated to $M$, to equal the dimension of the CDH).


To further illustrate this point,
in Fig. \ref{fig:polaron_truncation} we study the convergence of the average ground state magnetization at resonance for different choices of $M_{P}$. We conclude that 
the ultrastrong coupling limit up to $\lambda\approx2\Delta$ is well captured even for $M_P=M$. However, for the deep-strong coupling limit, $M_{P}$ has to be taken larger, here equal to $20$, the dimension required to converge the expectation value in the bare representation.


\subsection{Cavity observables: Photon number and squeezing}

We study here the performance of the 
CDH for predicting cavity observables. We compute the average photon number, $\langle \hat{a}^\dagger \hat{a} \rangle^{\text{eq}}$, and the squeezing term, $\langle \hat{a}^{\dagger2} + \hat{a}^2 \rangle^{\text{eq}}$ assuming the hybrid system is at a thermal state with $\beta=1$. 
We focus on the Rabi model, and consider a resonance scenario, where the two-level splitting $2\Delta$ equals the frequency of the cavity, $\Omega=2\Delta$. In the bare representation, we compute
\begin{equation}
    \begin{aligned}
        \langle \hat{\mathcal{O}} \rangle^{\text{eq}} = \text{Tr} \left( \frac{e^{-\beta \hat{H}}}{\mathcal{Z}}  \hat{\mathcal{O}} \right),
    \end{aligned}
\end{equation}
where $\mathcal{Z}=\text{Tr}(e^{-\beta \hat{H}})$ and $\hat{H}$ is the Rabi Hamiltonian for which we truncate the photonic Hilbert space to its $N$ lowest energy levels. For an order $M$ CDH we compute
\begin{equation}
    \begin{aligned}
        \langle \hat{\mathcal{O}} \rangle^{\text{eq}} = \text{Tr} \left( \frac{e^{-\beta \hat{H}^{\text{cdh}}_{(M)}}}{\mathcal{Z}} (\hat{U}_{P}^\dagger \hat{\mathcal{O}} \hat{U}_P)_{(M)} \right),
    \end{aligned}
\end{equation}
where $\mathcal{Z}=\text{Tr}(e^{-\beta \hat{H}^{\text{cdh}}_{(M)}})$, and the polaron rotation is done numerically while keeping the $M_P \geq M$ lowest energy levels, before truncating to the final $M$ levels. 


In Fig. \ref{fig:cavity} we find the CDH results (dashed) to converge faster to the numerically exact solution (full purple line) than the bare representation (full) up to the ultrastrong coupling (USC) regime. The squeezing term is zero for $N$, $ M \leq 2$. These simulations demonstrate that the CDH shows an advantage over the bare approach up to ultra-strong coupling when calculating the number operator. However, no such advantage is observed for squeezing. 

We further examine the convergence of the number operator and the squeezing when using low order CDH ($M=2$), while increasing the number of levels used in the mapping stage, $M_P$. As shown in Fig. \ref{fig:cavity}, the accuracy of the CDH is determined by $M_P$. This is encouraging: it means that we can reconstruct cavity observables at a much lower computational cost than in the bare representation. 

\begin{figure}[hbpt]
\centering
\includegraphics[width=0.9\textwidth]{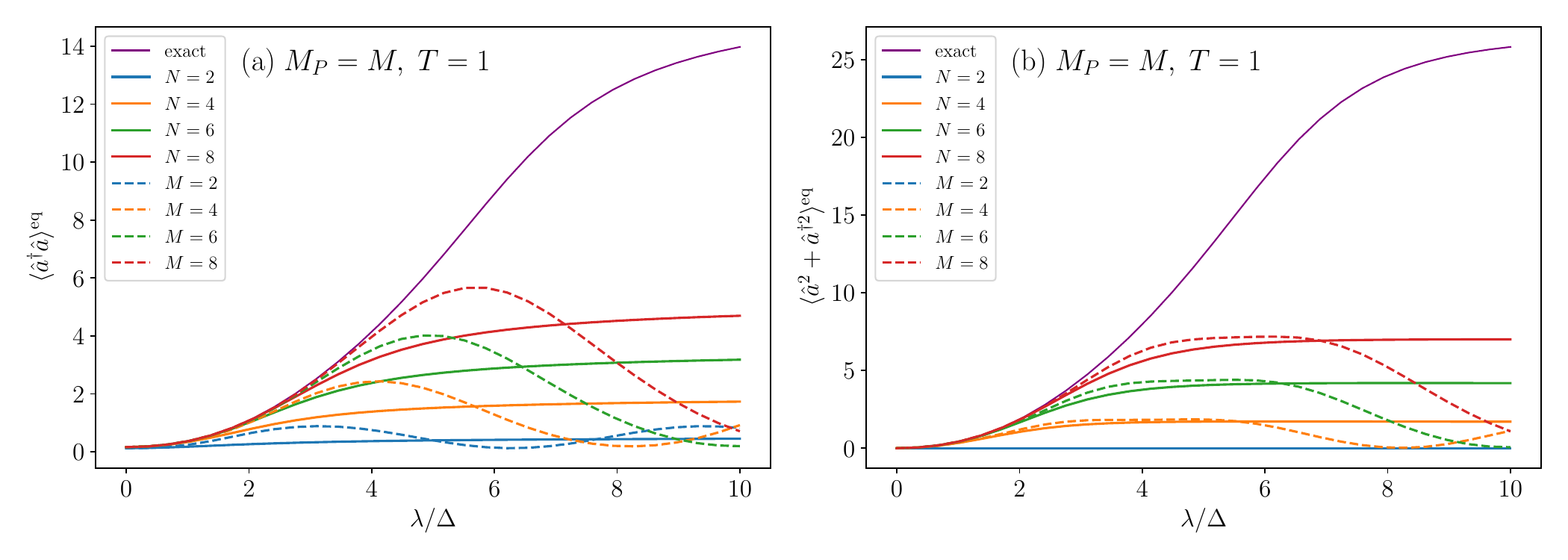}
\caption{ Average photon number and squeezing term with respect to the normalized coupling strength. Colored lines correspond to varying $N$ and $M$, with $M_P=M$. Other parameters are $\Omega=2$, $\Delta=1$, $\beta=1$. }
\label{fig:cavity}
\end{figure}

\begin{figure}[hbpt]
\centering
\includegraphics[width=0.9\textwidth]{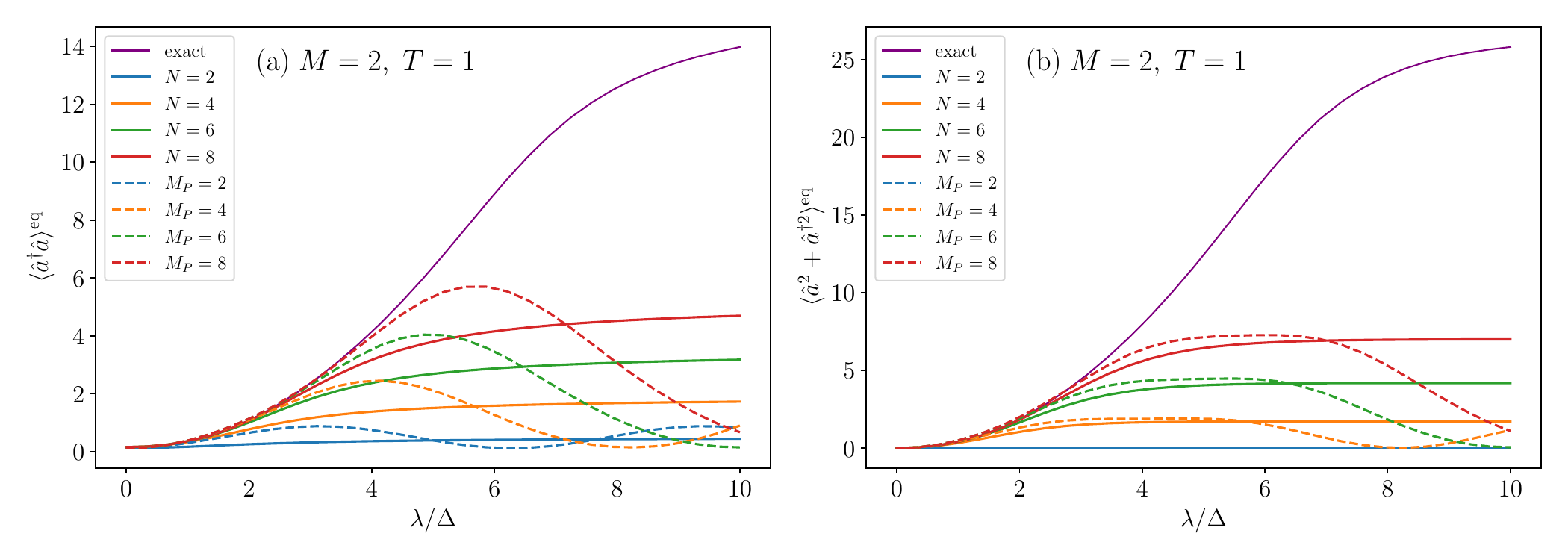}
\caption{ Average photon number and squeezing term with respect to the normalized coupling strength. Colored lines correspond to varying $N$ and $M_P$, with $M=2$. Other parameters are $\Omega=2$, $\Delta=1$, $\beta=1$. }
\label{fig:cavity2}
\end{figure}

\subsection{Dissipative dynamics of the Rabi model}

The CDH formalism allows us to study any systems coupled to a bosonic bath, described by a Hamiltonian $\hat{H}_O = \hat{H} +\hat{H}_B + \hat{S}\otimes\hat{B}$, where $\hat{H}_B = \sum_k \omega_k\hat{b}_{k}^{\dagger}\hat{b}_{k}$ and $\hat{B} = \sum_k f_k (\hat{b}_{k}^{\dagger}+\hat{b}_{k})$. For the mapping, see Sec. \ref{sec:Abaths}. In what follows, we focus on the resulting open system dynamics.

Under the Markovian and weak system-bath approximation, the dynamics of the open system is described by a quantum master equation $\dot{\rho}= \mathcal{L}\rho$, where we choose the dynamics generator $\mathcal{L}$ to be the Redfield tensor \cite{NitzanBook}. 
We assume the bath spectral density to be Ohmic, $J(\omega) = \sum_k |f_k|^2 \delta(\omega_k-\omega)= \gamma \omega e^{- |\omega|/\omega_c}$, where $\gamma$ is a dimensionless coupling parameter and $\omega_c$ is a frequency cutoff. For sufficiently small $\gamma$, the steady state of $\mathcal{L}$ is the Gibbs state. 
One could consider scenarios where coupling to the thermal bath is strong, which would be treated with methods such as reaction coordinate extraction \cite{PhysRevA.104.052617}, but we consider that out of scope of this work. For concreteness, we assume $\gamma=0.01$. As is commonly done in Redfield calculations, we ignore the lamb shift.

In the Main part of the paper, we presented in Fig.2(e)-(f)  the eigenvalues of the dissipative generator using the bare representation of the Rabi model and the CDH. We found that the spectrum of the generator  with the CDH converged for small $M$ for $\lambda$ up to the USC regime, unlike the case of the bare representation with $N=3$. Furthermore, the CDH framework provided physical insights into the structure of the eigenvalues of the generator.

\begin{figure}[hbpt]
\centering
\includegraphics[width=0.9\textwidth]{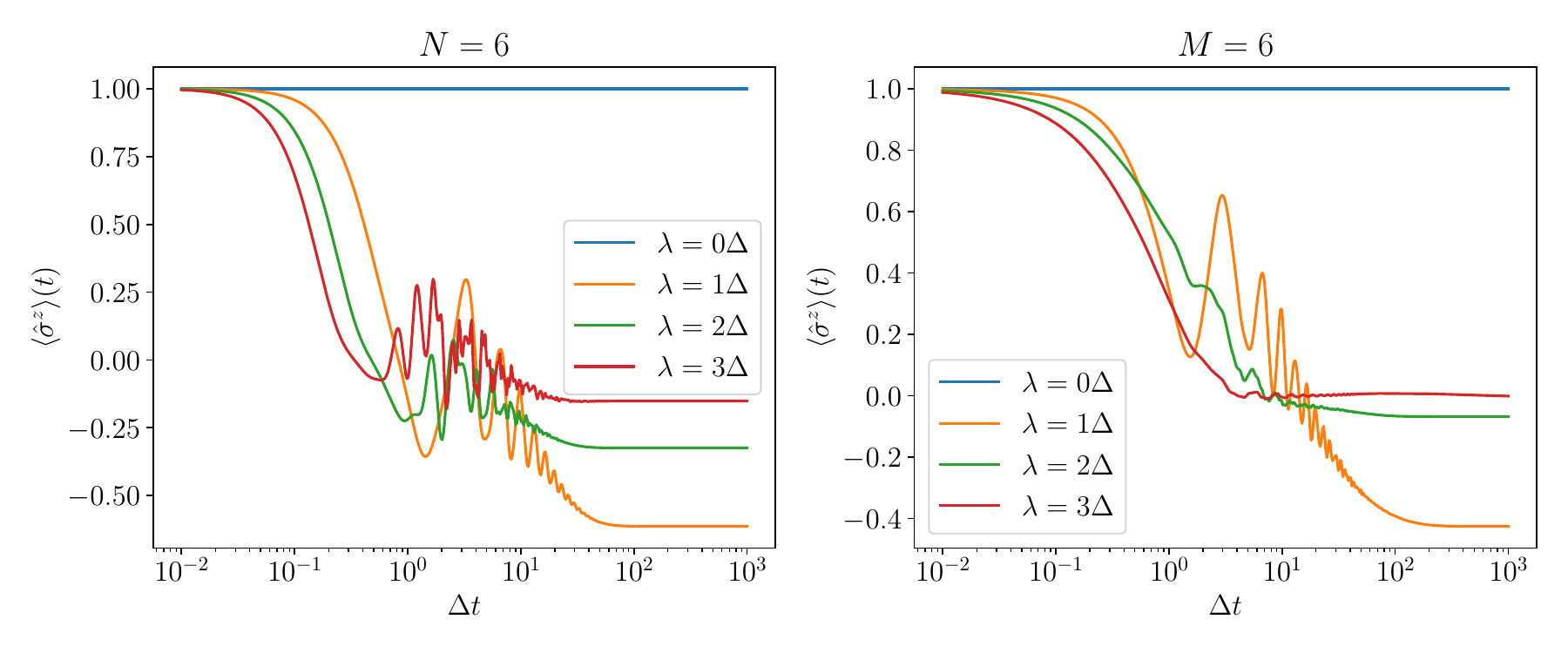}
\caption{Dynamics of the magnetization $\langle \hat{\sigma}^z\rangle(t)$ of the open Rabi model for varying $\lambda$. Numerical convergence was ensured with $M=6$ (right) for the CDH, while the bare representation with $N=6$ (left) is inaccurate in the given range of $\lambda$. We use $\Omega=2$ and $\Delta=1$ for the cavity and system's splitting. The dissipative bath, coupled to the cavity, is described by an ohmic spectral function with high frequency cutoff and a dimensionless coupling $\gamma=0.01$.
The spin system was initialized in its excited state, and the cavity in a thermal state with $T=1\Delta$. The temperature of the dissipative bath is $T=1\Delta.$}
\label{fig:Fig_4_resub}
\end{figure}

To supplement results in Fig.2(e)-(f) in the main text, we present in Fig. \ref{fig:Fig_4_resub} the magnetization dynamics for the open Rabi model at different coupling strengths. 
Using the CDH framework (right, $M$=6), we find that as we increase $\lambda$ towards the ultra-strong coupling regime, oscillations in the intermediate time range decay faster, {\it without changing their frequency}. The effect of oscillations suppression at strong coupling is due to eigenvalues of the dynamical generator moving in the negative real direction, see Fig.2(f) in the main text. The frequencies being fixed with $\lambda$  is uncovered in the branch structure that remain fixed with $\lambda$ at strong coupling.
%
In contrast, the bare representation with $N=6$, Fig. \ref{fig:Fig_4_resub} (left)  
inaccurately predicts both the dynamics and the steady state of the magnetization. In Fig.2(e) in the main text we recognize that the inaccuracy in the oscillatory behavior under the bare representation stems from eigenvalues whose imaginary part grows (diverges) with $\lambda$, rather than staying fixed.



\section{Dicke-Heisenberg model}
\label{sec:AppC}

\subsection{CDH Mapping}
We provide here details on the derivation of the CDH of the Dicke-Heisenberg model, illustrated for $M=3$. 
The model Hamiltonian is given by Eq. (7) in the main text, and we map it for general choice of lattice parameters. 
Its CDH form is given by the general result, Eq. (8) in the main text. It includes the transformed system (lattice chain), an all-to-all interaction 
emerging from $(\hat S^2)^{\rm{cdh}}_{(M)}$, where
 $\hat S= \sum_i{\hat \sigma_i^x}$, and a diagonal contribution from the cavity. 

We start by pointing out that the terms of one-body in Eq. (8) in the main text have a structure identical to that of the quantum Rabi model. 
Let us set $\epsilon=\lambda/(\Omega\sqrt{L})$. The two-body mapped system operators takes the form
\begin{equation}
\begin{aligned}\left(\hat{\sigma}_{i}^{x}\hat{\sigma}_{i+1}^{x}\right)_{(3)}^{\text{cdh}} & =\begin{pmatrix}\hat{\sigma}_{i}^{x}\hat{\sigma}_{i+1}^{x} & 0 & 0\\
0 & \hat{\sigma}_{i}^{x}\hat{\sigma}_{i+1}^{x} & 0\\
0 & 0 & \hat{\sigma}_{i}^{x}\hat{\sigma}_{i+1}^{x}
\end{pmatrix},\\
\left(\hat{\sigma}_{i}^{y}\hat{\sigma}_{i+1}^{y}\right)_{(3)}^{\text{cdh}} & =\begin{pmatrix}f_{0}(\epsilon)\hat{\sigma}_{i}^{y}\hat{\sigma}_{i+1}^{y}+g_{0}(\epsilon)\hat{\sigma}_{i}^{z}\hat{\sigma}_{i+1}^{z} & -ih(\epsilon)\left(\hat{\sigma}_{i}^{y}\hat{\sigma}_{i+1}^{z}+\hat{\sigma}_{i}^{z}\hat{\sigma}_{i+1}^{y}\right) & w(\epsilon)\left(\hat{\sigma}_{i}^{z}\hat{\sigma}_{i+1}^{z}-\hat{\sigma}_{i}^{y}\hat{\sigma}_{i+1}^{y}\right)\\
ih(\epsilon)\left(\hat{\sigma}_{i}^{y}\hat{\sigma}_{i+1}^{z}+\hat{\sigma}_{i}^{z}\hat{\sigma}_{i+1}^{y}\right) & f_{1}(\epsilon)\hat{\sigma}_{i}^{y}\hat{\sigma}_{i+1}^{y}+g_{1}(\epsilon)\hat{\sigma}_{i}^{z}\hat{\sigma}_{i+1}^{z} & iv(\epsilon)\left(\hat{\sigma}_{i}^{y}\hat{\sigma}_{i+1}^{z}+\hat{\sigma}_{i}^{z}\hat{\sigma}_{i+1}^{y}\right)\\
w(\epsilon)\left(\hat{\sigma}_{i}^{z}\hat{\sigma}_{i+1}^{z}-\hat{\sigma}_{i}^{y}\hat{\sigma}_{i+1}^{y}\right) & -iv(\epsilon)\left(\hat{\sigma}_{i}^{y}\hat{\sigma}_{i+1}^{z}+\hat{\sigma}_{i}^{z}\hat{\sigma}_{i+1}^{y}\right) & f_{2}(\epsilon)\hat{\sigma}_{i}^{y}\hat{\sigma}_{i+1}^{y}+g_{2}(\epsilon)\hat{\sigma}_{i}^{z}\hat{\sigma}_{i+1}^{z}
\end{pmatrix},\\
\left(\hat{\sigma}_{i}^{z}\hat{\sigma}_{i+1}^{z}\right)_{(3)}^{\text{cdh}} & =\begin{pmatrix}f_{0}(\epsilon)\hat{\sigma}_{i}^{z}\hat{\sigma}_{i+1}^{z}+g_{0}(\epsilon)\hat{\sigma}_{i}^{y}\hat{\sigma}_{i+1}^{y} & ih(\epsilon)\left(\hat{\sigma}_{i}^{y}\hat{\sigma}_{i+1}^{z}+\hat{\sigma}_{i}^{z}\hat{\sigma}_{i+1}^{y}\right) & -w(\epsilon)\left(\hat{\sigma}_{i}^{z}\hat{\sigma}_{i+1}^{z}-\hat{\sigma}_{i}^{y}\hat{\sigma}_{i+1}^{y}\right)\\
-ih(\epsilon)\left(\hat{\sigma}_{i}^{y}\hat{\sigma}_{i+1}^{z}+\hat{\sigma}_{i}^{z}\hat{\sigma}_{i+1}^{y}\right) & f_{1}(\epsilon)\hat{\sigma}_{i}^{z}\hat{\sigma}_{i+1}^{z}+g_{1}(\epsilon)\hat{\sigma}_{i}^{y}\hat{\sigma}_{i+1}^{y} & -iv(\epsilon)\left(\hat{\sigma}_{i}^{y}\hat{\sigma}_{i+1}^{z}+\hat{\sigma}_{i}^{z}\hat{\sigma}_{i+1}^{y}\right)\\
-w(\epsilon)\left(\hat{\sigma}_{i}^{z}\hat{\sigma}_{i+1}^{z}-\hat{\sigma}_{i}^{y}\hat{\sigma}_{i+1}^{y}\right) & iv(\epsilon)\left(\hat{\sigma}_{i}^{y}\hat{\sigma}_{i+1}^{z}+\hat{\sigma}_{i}^{z}\hat{\sigma}_{i+1}^{y}\right) & f_{2}(\epsilon)\hat{\sigma}_{i}^{z}\hat{\sigma}_{i+1}^{z}+g_{2}(\epsilon)\hat{\sigma}_{i}^{y}\hat{\sigma}_{i+1}^{y}
\end{pmatrix}.
\end{aligned}
\end{equation}
The dressing functions are given by
\begin{equation}
\begin{aligned}f_{0}(\epsilon) & =\frac{1}{2}\left(1+e^{-8\epsilon^{2}}\right),\\
g_{0}(\epsilon) & =\frac{1}{2}\left(1-e^{-8\epsilon^{2}}\right),\\
f_{1}(\epsilon) & =\frac{1}{2}\left(1+e^{-8\epsilon^{2}}\left(1-16\epsilon^{2}\right)\right),\\
g_{1}(\epsilon) & =\frac{1}{2}\left(1-e^{-8\epsilon^{2}}(1-16\epsilon^{2})\right),\\
f_{2}(\epsilon) & =\frac{1}{2}\left(1+ e^{-8\epsilon^{2}}\left(1-32\epsilon^2+128\epsilon^{4}\right)\right),\\
g_{2}(\epsilon) & =\frac{1}{2}\left(1-e^{-8\epsilon^{2}}\left(1-32\epsilon^2+128\epsilon^{4}\right)\right),\\
h(\epsilon) & =2\epsilon e^{-8\epsilon^{2}},\\
v(\epsilon) & =-2\sqrt{2}e^{-8\epsilon^{2}}\epsilon\left(8\epsilon^{2}-1\right),\\
w(\epsilon) & =4\sqrt{2}e^{-8\epsilon^{2}}\epsilon^{2}.
\end{aligned}
\end{equation}
Note that if $\gamma_{z}=\gamma_{y}$, the two body term in the Hamiltonian, emerging from the Heisenberg interactions, become block-diagonal.

Putting together all terms, the CDH Hamiltonian of the Dicke-Heisenberg model is explicitly given by
\begin{equation}
\begin{aligned}
(\hat{H})_{00}^{\text{cdh}} &= \sum_{i=1}^{L} \Bigl(
\tilde{\Delta}\hat{\sigma}_{i}^{z}
+ \gamma_{x}\hat{\sigma}_{i}^{x}\hat{\sigma}_{i+1}^{x}
+ [\gamma_{y}f_{0}+\gamma_{z}g_{0}]\hat{\sigma}_{i}^{y}\hat{\sigma}_{i+1}^{y}
+ [\gamma_{y}g_{0}+\gamma_{z}f_{0}]\hat{\sigma}_{i}^{z}\hat{\sigma}_{i+1}^{z}
\Bigr)
- \frac{\lambda^{2}}{\Omega L}\sum_{i=1}^{L}\sum_{j=1}^{L}\hat{\sigma}_{i}^{x}\hat{\sigma}_{j}^{x},\\[6pt]
(\hat{H})_{11}^{\text{cdh}} &= \sum_{i=1}^{L} \Bigl(
\tilde{\Delta}(1-4\epsilon^{2})\hat{\sigma}_{i}^{z}
+ \gamma_{x}\hat{\sigma}_{i}^{x}\hat{\sigma}_{i+1}^{x}
+ [\gamma_{y}f_{1}+\gamma_{z}g_{1}]\hat{\sigma}_{i}^{y}\hat{\sigma}_{i+1}^{y}
+ [\gamma_{y}g_{1}+\gamma_{z}f_{1}]\hat{\sigma}_{i}^{z}\hat{\sigma}_{i+1}^{z}
\Bigr)
- \frac{\lambda^{2}}{\Omega L}\sum_{i=1}^{L}\sum_{j=1}^{L}\hat{\sigma}_{i}^{x}\hat{\sigma}_{j}^{x}
+ \Omega,\\[6pt]
(\hat{H})_{22}^{\text{cdh}} &= \sum_{i=1}^{L} \Bigl(
\tilde{\Delta}(8\epsilon^{4}-8\epsilon^{2}+1)\hat{\sigma}_{i}^{z}
+ \gamma_{x}\hat{\sigma}_{i}^{x}\hat{\sigma}_{i+1}^{x}
+ [\gamma_{y}f_{2}+\gamma_{z}g_{2}]\hat{\sigma}_{i}^{y}\hat{\sigma}_{i+1}^{y}
+ [\gamma_{y}g_{2}+\gamma_{z}f_{2}]\hat{\sigma}_{i}^{z}\hat{\sigma}_{i+1}^{z}
\Bigr)
- \frac{\lambda^{2}}{\Omega L}\sum_{i=1}^{L}\sum_{j=1}^{L}\hat{\sigma}_{i}^{x}\hat{\sigma}_{j}^{x}
+ 2\Omega,\\[6pt]
(\hat{H})_{01}^{\text{cdh}} &= \sum_{i=1}^{L} \Bigl(
2 i \epsilon \tilde{\Delta}\hat{\sigma}_{i}^{y}
+ i h (\gamma_{z}-\gamma_{y})[\hat{\sigma}_{i}^{y}\hat{\sigma}_{i+1}^{z}+\hat{\sigma}_{i}^{z}\hat{\sigma}_{i+1}^{y}]
\Bigr),\\[6pt]
(\hat{H})_{02}^{\text{cdh}} &= \sum_{i=1}^{L} \Bigl(
2\sqrt{2}\epsilon^{2}\tilde{\Delta}\hat{\sigma}_{i}^{z}
+ w(\gamma_{y}-\gamma_{z})[\hat{\sigma}_{i}^{z}\hat{\sigma}_{i+1}^{z}-\hat{\sigma}_{i}^{y}\hat{\sigma}_{i+1}^{y}]
\Bigr),\\[6pt]
(\hat{H})_{12}^{\text{cdh}} &= \sum_{i=1}^{L} \Bigl(
-2\sqrt{2}\,i\,\epsilon(2\epsilon^{2}-1)\tilde{\Delta}\hat{\sigma}_{i}^{y}
+ i v (\gamma_{y}-\gamma_{z})[\hat{\sigma}_{i}^{y}\hat{\sigma}_{i+1}^{z}+\hat{\sigma}_{i}^{z}\hat{\sigma}_{i+1}^{y}]
\Bigr).
\end{aligned}
\end{equation}
Here, we provide expressions for a CDH of order 3.
These expressions are valid for a general anisotropy, $\gamma_x$, $\gamma_y$ and $\gamma_z$. 

We diagonalize this Hamiltonian and find the ground state of the Dicke-Heisenberg Hamiltonian. We then compute corresponding phase diagrams for different values of the Heisenberg coupling constants $\gamma_\alpha$. In the deep-strong coupling limit, $\lambda \rightarrow \infty$ the Hamiltonian becomes
\begin{equation}
\begin{aligned}
(\hat{H})_{00}^{\text{cdh}} &= \sum_{i=1}^{L} \Bigl(
\gamma_{x}\hat{\sigma}_{i}^{x}\hat{\sigma}_{i+1}^{x}
+ \frac{\gamma_y+\gamma_z}{2}[\hat{\sigma}_{i}^{y}\hat{\sigma}_{i+1}^{y}
+ \hat{\sigma}_{i}^{z}\hat{\sigma}_{i+1}^{z}]
\Bigr)
- \lim_{\lambda\rightarrow\infty}\frac{\lambda^{2}}{\Omega L}\sum_{i=1}^{L}\sum_{j=1}^{L}\hat{\sigma}_{i}^{x}\hat{\sigma}_{j}^{x},\\[6pt]
(\hat{H})_{11}^{\text{cdh}} &= \sum_{i=1}^{L} \Bigl(
\gamma_{x}\hat{\sigma}_{i}^{x}\hat{\sigma}_{i+1}^{x}
+ \frac{\gamma_y+\gamma_z}{2}[\hat{\sigma}_{i}^{y}\hat{\sigma}_{i+1}^{y}
+ \hat{\sigma}_{i}^{z}\hat{\sigma}_{i+1}^{z}]
\Bigr)
-\lim_{\lambda\rightarrow\infty} \frac{\lambda^{2}}{\Omega L}\sum_{i=1}^{L}\sum_{j=1}^{L}\hat{\sigma}_{i}^{x}\hat{\sigma}_{j}^{x}
+ \Omega,\\[6pt]
(\hat{H})_{22}^{\text{cdh}} &= \sum_{i=1}^{L} \Bigl(
\gamma_{x}\hat{\sigma}_{i}^{x}\hat{\sigma}_{i+1}^{x}
+ \frac{\gamma_y+\gamma_z}{2}[\hat{\sigma}_{i}^{y}\hat{\sigma}_{i+1}^{y}
+ \hat{\sigma}_{i}^{z}\hat{\sigma}_{i+1}^{z}]
\Bigr)
- \lim_{\lambda\rightarrow\infty}\frac{\lambda^{2}}{\Omega L}\sum_{i=1}^{L}\sum_{j=1}^{L}\hat{\sigma}_{i}^{x}\hat{\sigma}_{j}^{x}
+ 2\Omega,\\[6pt]
(\hat{H})_{01}^{\text{cdh}} &= (\hat{H})_{02}^{\text{cdh}} = (\hat{H})_{12}^{\text{cdh}} = 0. \\[6pt]
\end{aligned}
\end{equation}

We provide next additional 
results of Dicke-Heisenberg model, testing different variant models and discussing convergence in models details.


\subsection{Dicke-XX Heisenberg model}

\subsubsection{Entanglement entropy and Magnetization}

In Fig. 3 in the main text we study the Dicke-XX Heisenberg chain, $\gamma_{x}=\gamma_{y}=\Omega/8,\;\gamma_{z}=0$ case.
%

First, we test convergence when increasing the CDH order, $M$. In Fig. \ref{fig:phase_diagrams_XX}(a)-(c) we present the average ground state magnetization $M_{z}$ as the order parameter for both the bare and the CDH representations with $N=20$ and $M=1$, $M=3$, respectively. 
We find that $M=1$ already captures key characteristics of the phase diagram, with details improving as we increase $M$.


\begin{figure*}[hbpt]
\centering \includegraphics[width=0.9\textwidth]{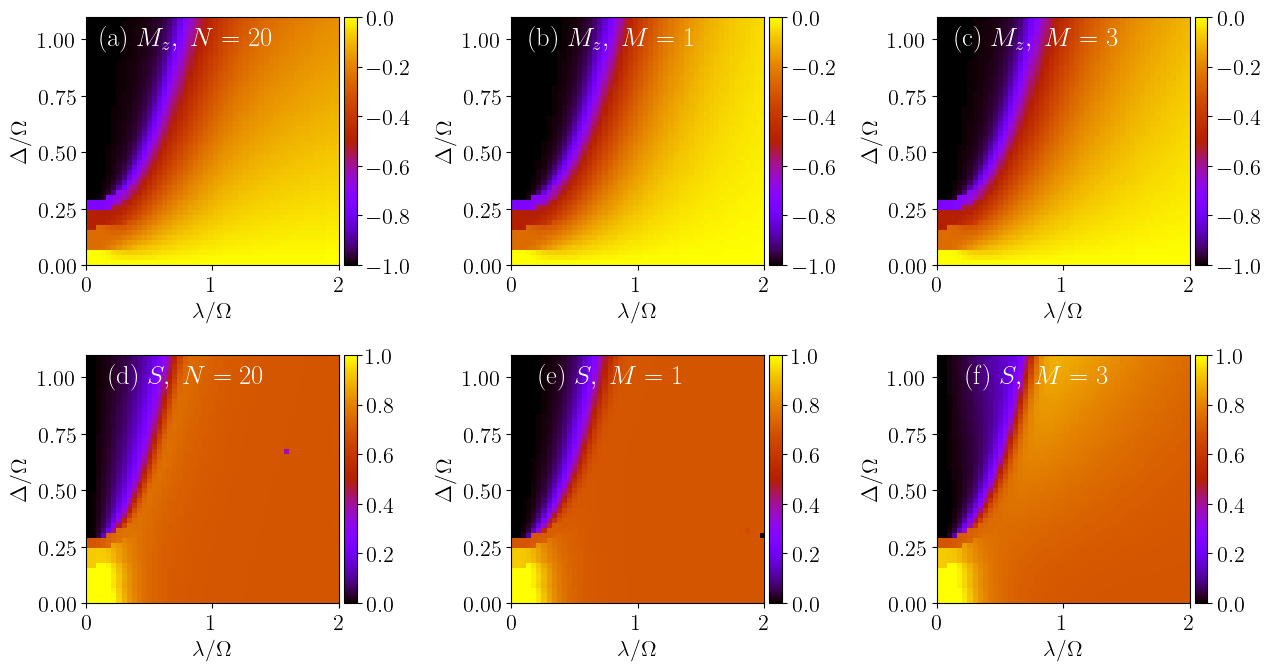} \caption{Phase diagrams of the Dicke-XX Heisenberg model using
the average magnetization $M_{z}$ (top) and the entanglement entropy $S$ (bottom) as order parameters. 
(a), (d) We perform simulations in the bare model and compare them to CDH results (b)-(c) with $M=1$ and (e)-(f) $M=3$. 
Model parameters are $L=8$, $\gamma_x=\gamma_y=\Omega/8$, ,$\gamma_z=0$, and $\Omega=2$.
}
\label{fig:phase_diagrams_XX} 
\end{figure*}

Second, we adopt a second order parameter: Fig. \ref{fig:phase_diagrams_XX}(d)-(e) shows results using the ground state entanglement entropy. We conclude that both magnetization and entanglement entropy measures faithfully reconstruct the phase diagram. 
%


\subsubsection{Size dependence}

Complementing Fig. 3 in the main text, we study in Fig. \ref{fig:structure_factors} the spin-spin correlation structure factor $S_{\alpha}=\sum_{i,j}\langle\hat{\sigma}_{i}^{\alpha}\hat{\sigma}_{j}^{\alpha}\rangle_{GS}/L^2$ of the ground state in the  Dicke-XX Heisenberg chain, where $i,j\in \{1, \cdots L \}$ and $\alpha\in \{x,y,z \}$. We observe that in the regime of weak cavity coupling, the correlations in the $z$ ($x$) direction are large (zero). When we increase the cavity coupling $\lambda$, the $z$ ($x$) correlations decrease (increase), as expected, due to the growing dominance of the all-to-all interaction term in the $x$ direction, generated by the cavity mode. 

We report that both $M=1$ and $M=3$ CDH models recreate the general trends of the structure factors, where the $M=3$ mapping gives accurate results for all values of $\lambda$. 
In Fig. \ref{fig:structure_factors} we further demonstrate that the absolute error of the CDH-computed structure factors, compared to the numerically exact solution 
gets progressively smaller for larger $L$. This observation extends a previous work \cite{Brett24}, where such an effect was observed for effective Hamiltonians corresponding to $M = 1$.

\begin{figure*}[hbpt]
\centering \includegraphics[width=0.9\textwidth]{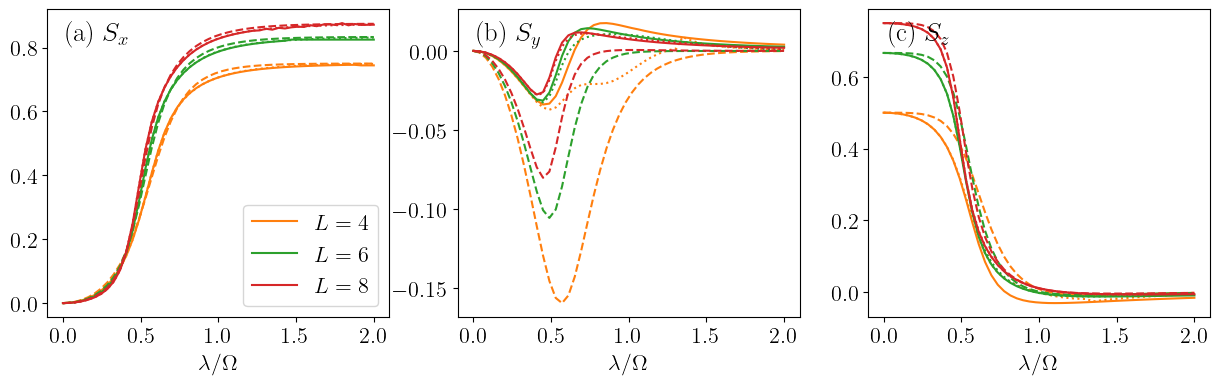}
\caption{Structure factors of the XX Dicke-Heisenberg chain with respect to
the scaled coupling strength $\lambda/\Omega$. Model parameters are $\Delta=\Omega/2$, $\gamma_x=\gamma_y=\Omega/8$, $\gamma_z=0$, and $\Omega=2$. Solid lines represent the numerically exact solution, dashed (dotted) lines correspond to CDH mapping results for $M=1$ ($M=3$). Line colors correspond to different chain lengths $L$.}
\label{fig:structure_factors} 
\end{figure*}

To investigate the behavior of the order parameters along lines of constant $\Delta$, we examine horizontal cuts of the magnetization in the Dicke–XX Heisenberg model, shown in Fig.~\ref{fig:phase_diagrams_XX}. Results for $\Delta / \Omega = 0.5$ are presented in Fig.~\ref{fig:PD_cuts_XX}. For all chain lengths, we observe that  magnetization curves cross at a specific point, indicative of a phase transition. The crossing point for a Dicke model ($\gamma_\alpha=0 \; \forall \alpha$) should occur at $\lambda/\Omega=0.5$, but the addition of the Heisenberg interaction here 
shifts that point, see inset of Fig. \ref{fig:PD_cuts_XX}.

\begin{figure*}[hbpt]
\centering \includegraphics[width=0.9\textwidth]{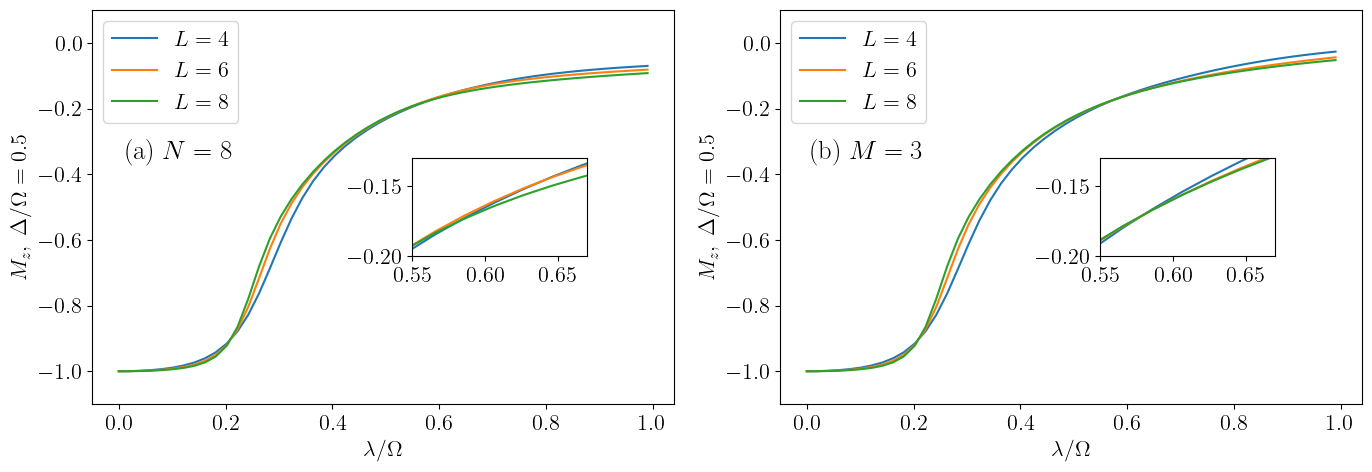} \caption{Cuts of the phase diagram of the  Dicke-XX Heisenberg model along the horizontal lines, where $\Delta/\Omega=0.5$. Results correspond to
the (a) bare model with $N=8$ and (b) the CDH with $M=3$. Insets zoom at the parameter regime
near the crossing points. Model parameters are $\gamma_x=\gamma_y=\Omega/8$, $\gamma_z=0$, and $\Omega=2$. We study different chains with lengths $L=4,6,8$.}
\label{fig:PD_cuts_XX} 
\end{figure*}

\subsection{Dicke-XXX Heisenberg model}

In Fig. \ref{fig:phase_diagrams_iso} 
we present the ground-state phase diagram of the {\it isotropic} Heisenberg model, also referred to as the Dicke-XXX Heisenberg model, where $\gamma_{x}=\gamma_{y}=\gamma_{z}=\Omega/6$.
Because to the symmetry of this XXX Heisenberg Hamiltonian, no transition from a ferromagnetic to a paramagnetic order is observed at weak coupling to the cavity.
This is because the $SU(2)$ symmetry is not broken by the Heisenberg interaction term. The phase diagram is consistent for both order parameters, magnetization and entanglement entropy. Both order parameters reveal the transition from an unentangled spin-ordered state to an entangled state with no magnetization as we increase the coupling strength.

\begin{figure*}[hbpt]
\centering \includegraphics[width=1\textwidth]{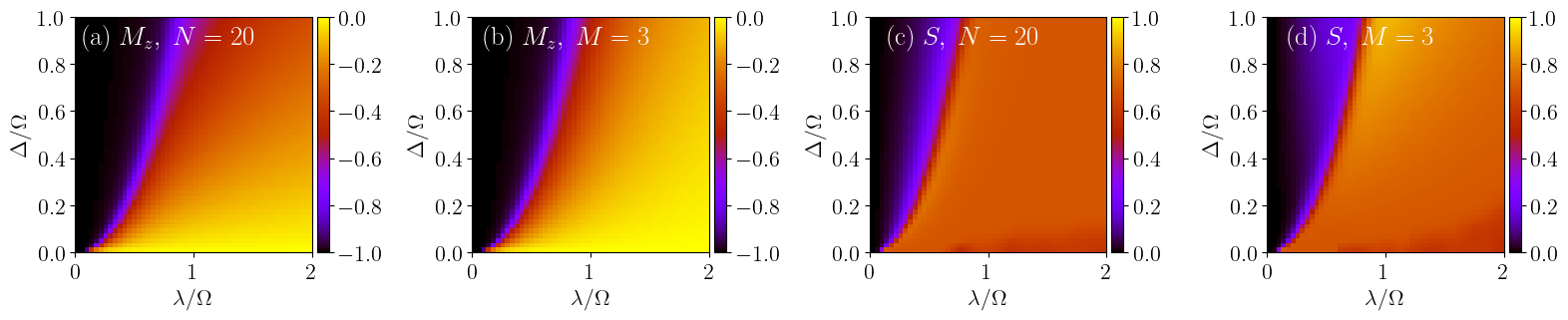} \caption{Phase diagrams of the  Dicke-XXX Heisenberg model. 
(a)-(b) The order parameter is the magnetization $M_{z}$ and we simulate it in both (a) bare and  (b) CDH representation.
(c)-(d) The order parameter is
the entanglement entropy $S$,
simulated in both (c) bare and  (d) CDH representations.
Model parameters are $L=8$ and $\gamma_x=\gamma_y=\gamma_z=\Omega/6,\; \Omega=2$.}
\label{fig:phase_diagrams_iso} 
\end{figure*}


\subsection{Dicke-Ising Model}

As a final validation of the CDH approach, we study the ground state phase diagram of the Dicke–Ising model with $L=8$. We set the parameters to $2\Delta = \Omega$ and $\gamma_x = \gamma_y = 0$, and reconstruct Fig.~3(b) from Ref.~\citenum{mendonca2025} in Fig.~\ref{fig:PD_Jachym}(a)-(b).  We identify three regions corresponding to the following phases in the thermodynamic limit:
antiferromagnetic-normal (yellow region), paramagnetic-superradiant (orange-red region), and ferromagnetic-normal (black region).
Both bare and CDH mapping faithfully reproduce the first- (left side) and second-order (right side) transitions between the normal and superradiant phases.

Closer inspection of the region between the antiferromagnetic-normal phase (yellow) and the paramagnetic-superradiant phase (orange-red) reveals signatures of another phase (discontinuities in $M_z$, marked with solid green arrows; dashed green arrows indicate discontinuities that do not prevail at larger $L$),
which we mark in Fig. \ref{fig:PD_Jachym}(c) for $L$=8. This phase was not reported in Ref.\citenum{mendonca2025}, probably due to insufficient resolution of the phase diagram computation, as explained in a recent comment \cite{hörmann2025commentrolematterinteractions}. This intermediate regime hosts an antiferromagnetic-superradiant phase, and was found and explained in Refs.\citenum{Zhang14,Schmidt2024,Schmidt2025,Kai25}. It is remarkable to note that CDH simulations of small lattices capture signatures of these phases, as we demonstrate by focusing on spin observables. 

\begin{figure*}[hbpt]
\centering \includegraphics[width=0.9\textwidth]{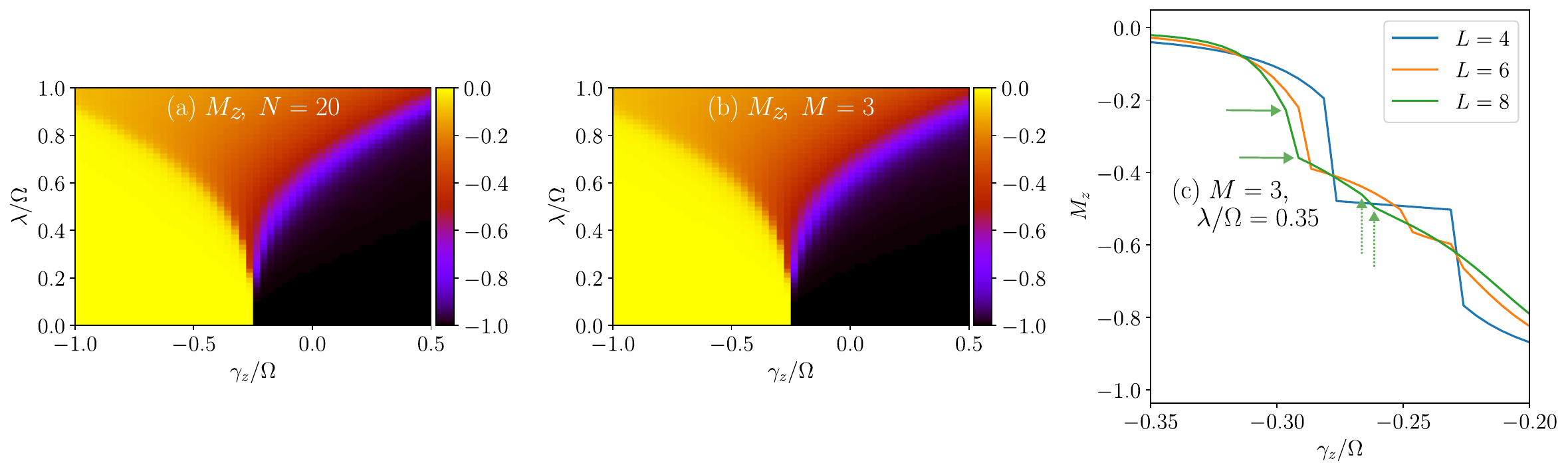}
\caption{Average magnetization $M_{z}$ for the Dicke-Ising model with $L=8$ using (a) bare Hamiltonian and (b) CDH mapping with $M=3$. (c) Cut of (b) along the line, where $\lambda/\Omega=0.35$ for different chain lengths. Heisenberg interaction parameters were $\gamma_{x} = \gamma_{y}=0$. Other parameters are $\Delta=1,\;\Omega=2$. Arrows highlight discontinuities in $M_z$ when $L=8$, with full arrows marking discontinuities that survive in the thermodynamic limit and dashed arrows marking discontinuities that disappear at large $L$ \cite{hörmann2025commentrolematterinteractions}.}
\label{fig:PD_Jachym} 
\end{figure*}

\end{widetext}

\bibliography{reference}

\end{document}